# High speed control of electro-mechanical transduction

## Advanced Drive Techniques for Optimized Step-and-Settle Response of MEMS Micromirrors

Matthias Imboden[1][1], Jackson Chang[1], Corey Pollock[2], Evan Lowell[2], Mehmet Akbulut[2], Jessica Morrison[3], Thomas Stark[4], Thomas G. Bifano[2], and David J. Bishop[1,2,3,4]

[1]Department of Electrical and Computer Engineering

[2]Department of Mechanical Engineering

[3]Department of Physics

[4]Division of Materials Science and Engineering

Boston University

Boston, MA 02215

May, 2016

*Micro/Nano Electro Mechanical Systems (MEMS/NEMS) provide the engineer with a powerful set of solutions to a wide variety of technical challenges. However, because they are mechanical systems, response times can be a limitation. In some situations, advanced engineered drive techniques can improve response times by as much as a thousand fold, significantly opening up the application space for MEMS/NEMS solutions.*

---

[1] Current affiliation LMTS, [École polytechnique fédérale de Lausanne](), rue de la Maladière 71b, 2002 Neuchâtel, Switzerland

MEMS are micromachines built using silicon micromachining processing techniques, similar to those found in Very-Large-Scale Integration fabs. Using a combination of patterning, deposition and etching techniques, one can build structures with microscopic moving parts with characteristic dimensions ranging from nanometers to millimeters. They differ from typical integrated circuits in their ability to leverage mechanical degrees of freedom to perform a function: They can be integrated with on-chip electronics to enhance their performance, optimizing the electromechanical transduction. Today, MEMS devices are a mature technology with a total market of $11.1B in 2014, expected to grow to over $20B by 2020. Applications range from airbag and pressure sensors in cars [1], microphones [2] and accelerometers in smart phones [3], [4], to micromirror displays [5]–[7]. MEMS devices fall into two broad categories: sensors and actuators. Sensors measure things like pressure [8], forces [9], [10] and acceleration, for example, to detect whether a car crash has occurred [11], [12]. Actuators are devices that move in response to a command. MEMS micromirrors, such as those shown in Figure 1, are examples of actuators that have the ability to steer and/or focus light by mechanically changing their shape and orientation. The focus of this article is how to steer MEMS actuators using advanced drive techniques. Such command shaping methods are used for rapid end point positioning, in this case with the aim of zero vibration. Fast point-to-point transitions are accomplished without actuating the resonant modes. While the focus is on micromirrors, the techniques are universally applicable to most actuator systems that can exhibit a resonant response.

MEMS micromirrors are an important subset of the actuator market. They are used in a wide range of applications to rapidly deflect and focus light. Figure 1 shows two such examples, both of which were fabricated using the multi-project wafer process PolyMUMPs by MEMSCAP [13]. Figure 1 a) is a commercial MEMS device is depicted with a mirror roughly 500 microns in diameter that can pivot about two axes using electrostatic actuation. An array of these mirrors form the optical switching element for the Lucent LambdaRouter [14]. Figure 1 b) shows a mirror driven by thermal bimorphs, which are composed of silicon and metals (in this case chromium and gold) that bend in response to a change in temperature. The small thermal capacity allows for low amounts of electrical power to generate significant temperatures in the bimorph structures. Optical beam steering for smart lighting systems is achieved by controlling the temperature of each of the four bimorphs. At the same time, the focus can be tuned by varying the temperature of the mirror itself [15]. MEMS micromirrors are used today for digital cameras [16], [17], network elements in optical networks [18], bar code scanners in supermarkets [19], retina scanners [20] and numerous other optical systems [21]–[23]. In all cases, an electrical signal is applied and the mirror responds, going from one set point to another. An important design attribute is the ability to do this quickly, i.e. a fast step-and-settle response. This article shows how using advanced drive techniques can reduce this step-and-settle time by a factor of over a thousand, a huge win for the systems designer.

The control methods discussed here allow the MEMS engineer to escape some of the constraints imposed by the physics of the response times of a simple harmonic oscillator. For example, when building a MEMS device such as a micromirror, there are tradeoffs between range of motion, response time and optical loss. In a standard micromirror system, the mirror is connected to a spring. Typically, a soft spring yields a larger range of motion than a stiff spring and, as the diameter of the mirror is increased, the optical losses decrease. Yet large mirrors are heavy, and combining them with soft springs turns them into high quality factor devices with low resonant frequencies, and consequently long settling times. However, the techniques discussed here can open up phase space significantly in terms of response times. For example, implemented on a commercial MEMS mirror used as an optical switch (discussed below), advanced control techniques can improve the step-and-settle response time from a relatively slow ~300 ms to just ~300 μs, an improvement of a factor of a thousand. Where slow devices have more space for improvement, it is also shown how the settling time of a much smaller and faster device is reduced from over 100 μs to just 17 μs. Given the scaling laws governing the dynamics, to achieve an improvement in response time by a factor of a thousand would require modifying either the mass or spring constant by a factor of a million. This is often impractical, if not impossible: decreasing the mass will diminish the optical efficiency and stiffening the springs impedes motion and angular range. Using advanced drive techniques eases the design constraints and helps make MEMS devices a better fit for a range of interesting and novel applications.

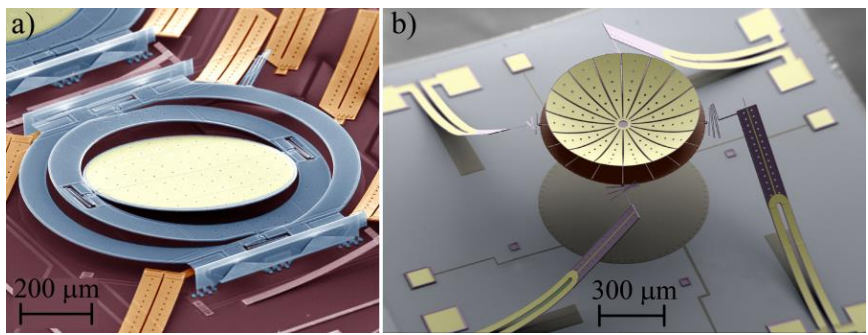

Figure 1. Two MEMS micromirrors examples. a) LambdaRouter, a commercial MEMS mirror built by Lucent Technologies. The mirror pivots around two axes and is used as the optical switching element in a large network cross-connect. Photo Courtesy of Lucent Technologies Inc. b) Varifocal mirror, driven by electrothermal bimorphs, used for beam-steering in a smart lighting application.

Many MEMS mirrors can be approximated as damped, driven harmonic oscillators. For such systems, the settling time is characterized by the ring-down time. After a mechanical perturbation, a linear dissipative system will lose amplitude exponentially as it oscillates (or settles) into a new static position. This settling time, $\tau$, defines the duration of the exponential decay and is a measure of the oscillation frequency ($\omega = \frac{2\pi}{T}$) and dissipation $Q^{-1}$ (the inverse of the quality factor)

$$\tau = \frac{2Q}{\omega} = \frac{Q}{\pi}\sqrt{\frac{m}{k}} = \frac{2m}{\gamma}, \quad (1)$$

where $m$ is the mass and $\gamma$ is the loss factor of the resonance mode. Relevant in the context described here is the relationship of the dissipation to the number of oscillations during the ring-down

$$N = \frac{\ln(2)}{\pi} Q. \quad (2)$$

$N$ is the number of times the resonator oscillates until losing ½ of its initial amplitude. It is noteworthy that the number of oscillations is independent of both the resonant frequency and the amplitude; this is only true for linear systems, such as those considered here.

Reducing the quality factor decreases the settling time, but a low $Q$, or high dissipation, is more consequential than just the equivalent of applying the brakes. The dissipation is a measure of the coupling of the resonant mode to the environment: the ability to remove energy from a resonator (through losses, i.e. low $Q$) is mirrored by the resonator's ability to sense its environment and absorb energy from the surroundings, often in the form of noise [24]. A major advantage of MEMS devices is the high frequency *and* concomitant low coupling to the environment, making them stable, mechanically quiet, sensitive and energy efficient. Both mechanical and electrical noise typically fall off as *1/f*, so operating at high frequencies pushes the dynamic response away from noise sources: a high quality factor decouples the mode from extrinsic disturbances and is therefore often desired. In certain MEMS designs, quality factors can exceed $10^6$ [25], [26] and frequencies can range from $10^2$ to $10^8$ Hz, producing transient times that can reach $10^4$ s (almost three hours!). This is impractical, as a useful device should have response times on the order of milli- or even microseconds. It is shown here, that by applying a specific drive force, termed a double-step drive, the ring-down can be completely eliminated, and force the MEMS device to settle in a time equal to half of its natural period of oscillation. The result is universal for such systems and has been implemented in wide range of resonant systems including MEMS [27]–[32]. While this article presents an analytical solution for linear (or almost linear) systems, input shaping for non-linear systems has also been extensively studied [33]–[36]. Provided that there are minimal drift or other instabilities, this method can replace more complex closed-loop systems which also offer rapid settling times [31], [37]–[39], but require sensing for feedback using PID (proportional/integral/differential), additional control electronics, or learning algorithms [40]. Even in cases where active feedback is required due to high precision requirements or creep in the MEMS material, utilizing the proposed double-step drive technique provides an excellent prediction of the response and can help stabilize the system. The double-step drive presented is well known in control theory, specifically as a feedforward, or input shaping approach [41]–[43]. In this article a time-domain derivation is used to determine the fastest possible settling time of a resonant system. The results are equivalent to solutions based on command shaping of second order systems [44], [45]. To aid readers familiar with the standard control theory formalism the well-known prefilter transfer functions are included for key results. A common application of feedforward drive schemes is in the stabilization of piezoelectrically driven atomic force microscope cantilevers. This type of MEMS is not described here but is well covered in the existing literature [42].

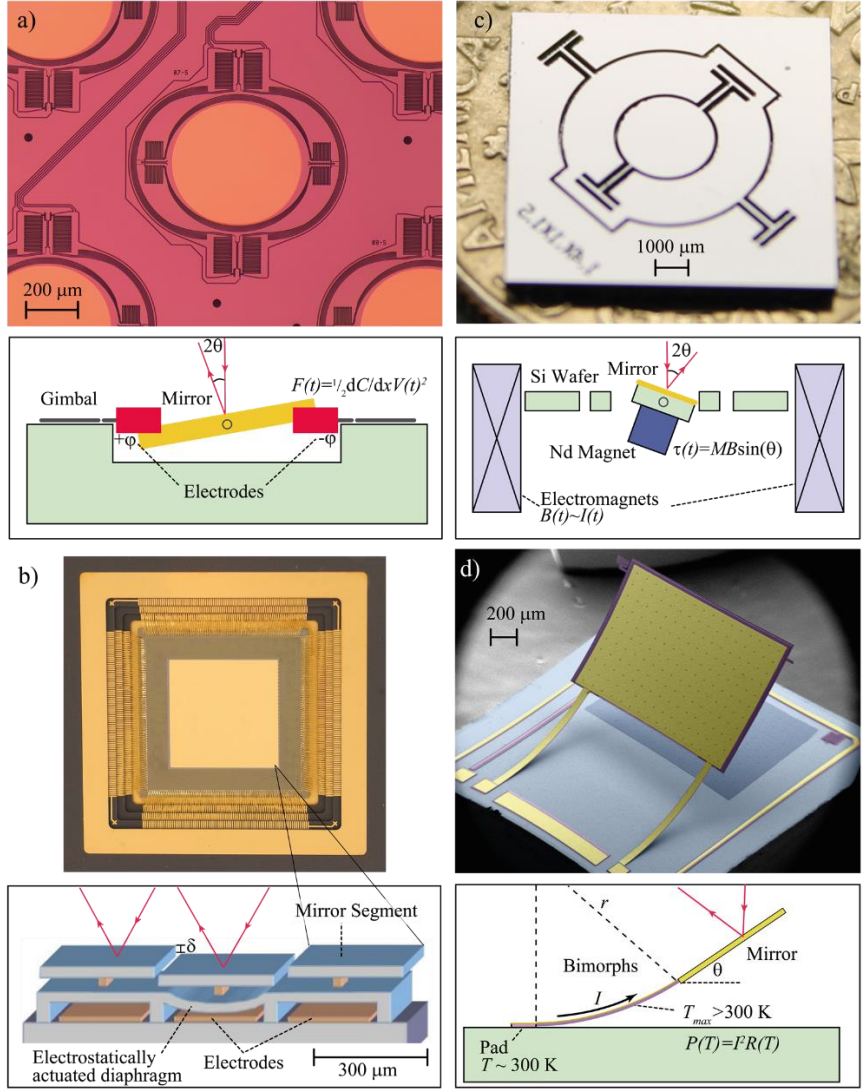

Figure 2. MEMS mirrors using different electromechanical transduction schemes. a) Optical micrograph of a commercial optical cross-connect mirror by CrossFiber. The electrostatically actuated MEMS uses a gimbal design. Four sets of capacitors are needed to generate the electrostatic bi-directional force along two axes. b) Array of mirror elements forming a Spatial Light Modulator. SLMs correct optical wavefront phase errors in imaging systems such as microscopes and telescopes. (adapted with permission from [27]) c) Gimbal design for two axis rotation, driven by a permanent magnet mounted to the central mirror (shown in lower diagram) and two sets of orthogonal electromagnets. d) Thermally driven single axis mirror. Joule heating results in strain gradients along the thickness of the bimorphs changing their radius of curvature, thus moving the mirror.

In the next section the analytic model is presented, showing how to eliminate ring-down by applying advanced drive forces to a linear damped resonator. The only prior knowledge required to apply this technique are the mode period and dissipation as well as the assumption that the device is at rest before any change in amplitude is attempted[†]. No sensing or active feedback is required; this greatly reduces complexity and computational power needed for controlled point-to-point electromechanical transduction. The practical application of this theory is demonstrated on the four devices illustrated in Figure 2. These MEMS devices use electrostatic, electromagnetic, or electrothermal forcing to deflect a beam of light. In each case, the settling time is significantly reduced, validating the universality of the model and demonstrating specific features of the approach. The examples provided are illustrative of the large achieved reductions of the settling time. In the final section, a method is considered involving overdriving the MEMS devices to further decrease the response time in a regime unconstrained by the intrinsic restoring forces of the device itself. This is a form of bang-bang control [46] which gives an optimum in settling time for a given maximum available force. Additional considerations are also presented on the effect of higher frequency modes, and an alternate resonant drive is presented. The robustness of the approach is also considered in detail, demonstrating the sensitivity to detuning of the drive parameters.

## Theory: The analytic solution to terminate ring-down

This section presents the theory governing the double-step drive technique. As will be shown, the point mass, linear low-dissipation model is sufficient to predict the precise settling time of many real world MEMS devices. Such a system is described by the well-known differential equation

$$m\ddot{x} + \gamma\dot{x} + kx = F(t), \quad (3)$$

where $x$ and the time derivatives thereof represents the position, speed and acceleration of a device of mass $m$, driven by the force $F(t)$ which is balanced by the restoring force of the spring characterized by the spring constant $k$. As the device moves, it loses energy at a rate proportional to the velocity, characterized by the loss factor $\gamma$, as introduced in the previous section. Provided with energy, a system described by (3) will oscillate if the damping is below the critical value of

$$\gamma < \gamma_c = 2m\omega_0. \quad (4)$$

The solution to this differential equation for a step input force ($F(t<0) = 0$ and $F(t>0) = F_0$), starting at rest from the origin ($x(0)=0$ m, $\dot{x}(0)=0$ m/s) is described in [47] and reproduced in the open-loop step response Sidebar

$$x(t) = \frac{F_0}{k}\left(1 - \frac{e^{-\frac{\gamma t}{2m}}}{\sin\phi}\sin\left(\sqrt{1-\left(\frac{\gamma}{2m\omega_0}\right)^2}\omega_0 t + \phi\right)\right), \quad (5)$$

with $\cos(\phi) = \frac{\gamma}{2m\omega_0}$. $F_0$ is the force required to reach and stay at the steady state position, $x_0$, after all ringing has subsided. Given the linear restoring force, this expression is valid for any change in forcing described by a step function for a MEMS mirror, initially at rest. In other words, valid for any transition between two set points resulting from two constant levels of applied force.

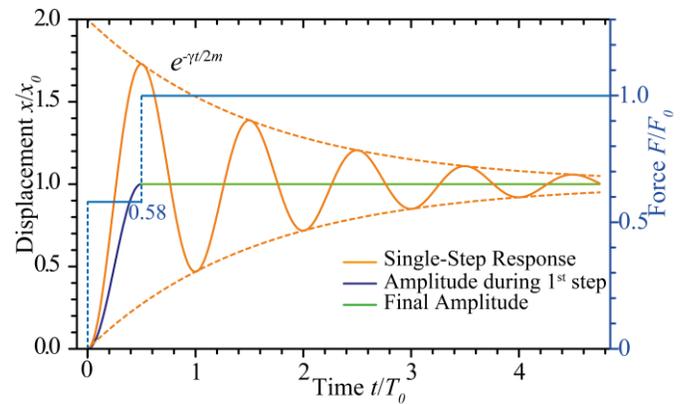

Figure 3. Single- and double-step response of a dissipative resonator. ($\gamma = 0.2$) results in the rather rapid decay of the single step response. The duration of the first step of the double-step forcing scheme is very close to $T_0/2$, while the forcing must be raised to $0.58 \times F_0$, where $F_0 = k\,x_0$ is the steady state force required to reach $x_0$.

---

[†] If not fully at rest then the device must be moving slowly compared to the typical actuation velocity i.e. ($v \ll \omega_0 x_0$). In this limit the device is quasi-static and the externally applied force is in balance with the restoring force.

Using equation (5), one can determine the force, $F_1$, and duration, $t_1$, needed to settle the oscillator without overshoot beyond a final resting position $x_0$. The requirements for such a system are a balance between the external forcing and spring restoring forces at position $x_0$ as well as a vanishing velocity at that position: If all forces are balanced and the system is at rest, then it will stay there indefinitely. The initial force needed will depend on the final desired position, the restoring spring constant and the dissipation. Essentially, one uses the overshoot of the dynamic response resulting from $F_1$ to reach the steady state when applying the force $F_0$. Normally the overshoot is an unwanted side effect of applying a step function drive to a resonator. Here, by carefully engineering the drive, the overshoot is leveraged to reach the desired position. Once the desired position is reached, the force is quickly switched from $F_1$ to $F_0$, completing the second step of the double-step drive. This results in the following two boundary conditions

$$x(t_1) = x_0$$
$$\dot{x}(t_1) = 0, \quad (6)$$

where $t_1$ is the time at which the resonator reaches the target position. Solving for the minimum non-vanishing value of $t_1$ at which time the system is at rest (as the system oscillates, there exist an infinite number of repeating moments of zero velocity), one obtains, using the second expression in equation (6)

$$t_1 = \frac{T_0}{2} \frac{1}{\sqrt{1-\left(\frac{\gamma}{2m\omega_0}\right)^2}}, \quad (7)$$

where $T_0$ is the period of oscillation of the actuated mode. By plugging $t_1$ into the first boundary condition in equation (5) and solving for $F_1$, one obtains an expression for the force required to reach $x_0$

$$F_1 = F_0 \left(1 - \frac{1}{1+e^{\frac{\gamma}{2m}t_1}}\right). \quad (8)$$

$F_1$, and $t_1$ are plotted as a function of the dissipation $\gamma/m$ in Figure 4. For vanishing dissipation $F_1 = F_0/2$ and $t_1 = T_0/2$. For a wide range of practical MEMS devices, this set of solutions works very well and is easy to implement in realistic drive circuits. As the dissipation approaches the critical value of $\gamma \to \gamma_c = 2m\omega_0$, $F_1 \to F_0$ and $t_1 \to \infty$. These two extremes are expected, as without dissipation the overshoot is twice the equilibrium value (due to the conservation of energy) and hence only half of the final force is required. Without dissipation, the maximum overshoot of the ringing occurs for the first time at half the period; hence the final position is reached at $t_1 = T_0/2$. In the high dissipation limit, no ringing occurs. No overshoot means that the full force must be applied and the device approaches the desired position asymptotically, taking a very long time to get there. Many MEMS devices, especially when operating in a vacuum, will be very close to the zero loss limit when implementing the double-step drive. Illustrated in Figure 4 b), in the limit $\gamma/(2m) \ll \omega_0$, the correction to $t_1$ is vanishingly small. The increase of the force grows linearly with dissipation but, as a practical matter, $F_1 > F_0/2$ must be considered only when driving MEMS devices with quality factors below roughly one hundred. The equivalent result can be obtained by considering the superposition of two ring-down response curves as described by equation (5), the first commencing at *t* = 0 and the second at *t* = *t₁*. As at *t₁* the system is at rest, the expression given in equation (5) is valid. For the correct initial forcing, the resulting ring-down curves add to a new constant position $x_0$ [27], [48]. Finite width pulses have also been considered as shaping inputs to minimize vibration, where, as in the case presented here, only the mode frequency and dissipation are required as input parameters [49]. The feedforward input sculpting produces the same solution as derived here, and is presented in detail in the Sidebar on Drive Filters. Important for control theory is knowledge of the pre-filter, which in this case is written as [44], [50], [51]

$$F(s) = F_1 + (1 - F_1)e^{-t_s s}. \qquad (9)$$

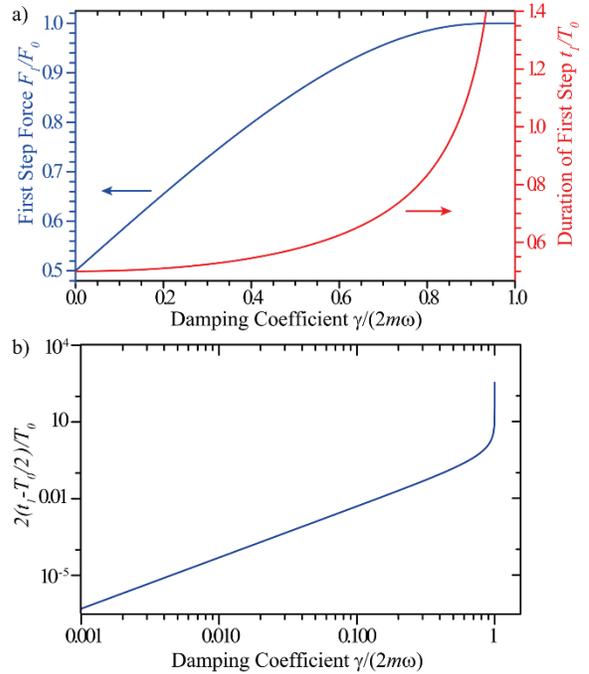

Figure 4. Force and timing of steps as a function of $\gamma/(2m\omega)$ a) Force and duration of 1st step in double-step actuation mode as a function of dissipation. For dissipation free systems the force and time are both half the final force and period respectively. With increasing dissipation the force increases linearly and saturates as the dissipation exceeds the critical limit. b) Log-log plot of the additional duration of the applied force $F_1$ beyond $T_0/2$ as a function of dissipation. The correction is only a small fraction of the oscillation period until the critical dissipation is approached at which point the time diverges.

The significance of this formalism, and how it helps control the vibrational modes of MEMS, is discussed for the two fold double-step drive of the magnetically driven mirror. This filter can be adapted to further reduce the settling time, in what is referred to as overdrive, and is demonstrated using finite element simulation later in this article. A variation has been experimentally implemented [31] and is discussed as well.

The calculated position as a function of time, along with the drive force, is plotted in Figure 3. The effect of dissipation on the initial force step is illustrated and, for comparison, the single-step response is included. The time axis is normalized to the period of the resonator. This drive method achieved the fastest possible settling time by eliminating all ringing, but the time is still determined by the mass and spring constant of the device because they set the fundamental resonant frequency of the undamped system and hence, the deceleration amplitude.

Naturally, a resonator can be accelerated almost arbitrarily quickly by increasing the initial drive force. The resulting speed will be too high for the restoring force of the spring to bring the device to rest when it reaches $x_0$. Consequently, a decelerating force must be applied. In the extreme forcing limit this approaches the particle in free space scenario. First one accelerates it and then decelerates it, arriving at its final position with no net force on it with zero velocity. This overdrive forcing method, or bang-bang control mode, along with an alternative resonant drive scheme is discussed later in this article. To conclude the stability of the double-step drive is analyzed with regards to forcing errors in time and amplitude. Fourier analysis is used to demonstrated how high frequency modes are affected by the various drive schemes and the effects of softening, or rounded the edges of the drive force are discussed (expected with any physical actuation).

In the next section, experimental implementations of the double-step drive are presented. It is shown that by using this drive scheme, the settling time of a commercial device can be reduced by three orders of magnitude.

## Experimental realization of the double-step drive for three MEMS transduction methods

The theory presented here is very general and applicable to all linear, harmonic systems. For this reason, the double-step drive has been implemented not only in mechanical systems, but any linear resonant system. The point mass equations described above are applicable to MEMS devices that can be characterized by a mode-dependent effective mass and spring constant. In this section, the double-step drive is illustrated on four different MEMS devices, using electrostatic, electromagnetic and electrothermal transduction. Each method has specific advantages and drawbacks. The technological details, specifically related to the control of the applied force, vary. Specifically, the drive signal may not be proportional to resulting force, and the force resulting from the drive signal may not result in a linear change of amplitude. The notation of single and double-step drive is with respect to the forcing. The actual drive signal, typically an applied voltage or current, may differ greatly from the step function *forces*. However, in all cases, a transfer function can be defined to adapt the drive and effectively translate it back to the linear equivalent. For example, as will be seen below for capacitive drives, the forcing is linear in *C* and hence for a linear spring system the amplitude (and force) is proportional to $V^2$. The ability to create this transfer function in a reliable way is often the limiting factor for implementing the double-step drive effectively in non-linear systems.

The electrostatic examples will illustrate the benefits of using the double-step drive on two commercial MEMS devices. For all measurements the location of a laser spot reflected of the micromirrors was measured using a position sensitive detector (PSD). A reduction in the settling time of three orders of magnitude is demonstrated for optical switches in a torsional system. The second capacitive example is that of a piston mode used for an SLM, which demonstrates an improved settling time by more than a factor of five. The high-quality factor magnetic scanning mirror example demonstrates that the double-step can be implemented even when multiple modes are driven by the applied step function in torque. Finally, considering a thermally actuated device, the difference between resonant actuation and creep is illustrated. The advanced drive scheme can facilitate rapid mirror positioning for devices with non-linear responses by separating the resonant and transient responses. The most relevant experimental results and parameters are tabulated in the Sidebar. It is worth noting that both the capacitively driven CrossFiber optical cross connect and the thermally actuated scanning mirror were controlled by an Arduino microprocessor and a simple amplifier circuit, at a cost of under $5 per channel. For each device the needed parameters, $f_n$ and $Q_n$, where determined experimentally.

**Electrostatic Drive: High speed, small angle deflections for low-power MEMS**

The electrostatic, or capacitive, force is one of the most common drive methods for MEMS. This is because the electrical signals are easy to generate and energy consumption is minimal, ideally only the charging energy of a very small capacitance. The resulting force $F_c$ is a function of a change in capacitance with respect to displacement

$$F_{C_i} = \frac{dE_C}{dx_i} = \frac{1}{2}\frac{dC}{dx_i}V^2, \qquad (10)$$

where $E_c$ is the energy of a capacitor of capacitance $C$ charged by the potential $V$, and the force acts along the dimension $i$. Because the force scales as $V^2$, the two capacitor plates are always attracted towards each other, no matter the polarity. To create bidirectional motion (or angles), two capacitors are used. The most common implementations include parallel plate or comb capacitors. Parallel plates are ideal for small displacements perpendicular to the capacitor plates, with $C_{pp} = \epsilon_0 \frac{A}{x_0-x} \approx \frac{\epsilon_0 A}{x_0}\left(1 + \frac{x}{x_0} + O[(\frac{x}{x_0})^2]\right)$ and $\frac{dC}{dx} \approx \frac{\epsilon_0 A}{x_0^2} + \cdots$. $A$ is the area, $\epsilon_0$ the permittivity of free space, and $x_0$ the capacitor gap when no voltage is applied. Examples include the LambdaRouter and Spatial Light Modulators (SLMs) depicted in Figure 1 and Figure 2 respectively. The above expression is valid only for the simplest case and must be modified for angular actuation where the capacitor plates are no longer parallel. The change in capacitance, and thereby the force, is inherently non-linear, both in displacement and drive voltage. Comb actuators are typically used for long-range linear displacement that is parallel to the capacitor, with a capacitance and force given by $C_{ca} = 2N\epsilon_0\frac{t\,x}{g}$ and $F_{ca} = N\epsilon_0\frac{t}{g}V^2$ respectively (the capacitance is often offset by an initial overlap of the combs). Here, $N$ is the number of combs, and $t$ and $g$ are the comb thickness and gap respectively. There are cases where a vertical comb capacitance configuration is used to generate a torque to pivot a mirror [52], [53] as is the case for the CrossFiber device discussed next.

The linearization of the force is straightforward for the comb actuator, where the drive parameter becomes $V^2$. The parallel plate capacitor is more complex. For small displacements, the force is again linear in $V^2$; however, for moderate displacements, this becomes a poor approximation. The force is a function of both voltage and the displacement. Therefore, when switching between two points, one must consider the instantaneous position during the crossover and adjust the applied voltage accordingly in order to maintain a constant force. As the displacement approaches the 1/3$^{rd}$ pull-in [54], the system becomes unstable and all higher terms must be considered. Reference [34] presents a model and simulation results for a non-linear multimodal MEMS with shaping control based on an iterative energy balance argument to extend the range of the double-step drive all the way to the pull-in point. Furthermore, it has been demonstrated that command-shaping methods can not only minimize motion induced vibration but also minimize impact in capacitive MEMS pull-in devices such as RF switches [36].

In the steady state, no current flows, so the power consumption is low. The electrical settling time is characterized by $\tau = RC$, $R$ being the lead resistance of the capacitor. As the capacitance is typically small (picofarads), $\tau$ is often on the order of nanoseconds, sufficiently fast for MEMS. The small capacitance is also one of the drawbacks of electrostatic MEMS, as it produces low driving forces. For parallel plate capacitors, the gap may be decreased to improve the electromechanical transduction; however, the resulting range falls off as quickly. To compensate for these effects, large gap devices require high voltages to operate, increasing the complexity of the drive circuits. Consequently, scanning mirrors often use comb capacitors which are not limited by the 1/3$^{rd}$ pull-in effect. Spatial Light Modulators which typically require a small throw, of order half the wavelength of the reflected light, usually use parallel plate capacitors as a drive mechanism.

The commercial optical cross connect device from CrossFiber shown in Figure 2 a) uses two sets of vertical capacitive comb drives and a gimbal to pivot a mirror about two axes. As it is proprietary, the technical specifications and details of this device will not be discussed here. It is straightforward, however, to measure the electromechanical transfer function and ring-down. All the parameters needed for a double-step drive can be extracted from the ring-down plot (Figure 5 a)). After measuring the mode frequency and corresponding quality factor, the double-step drive was implemented. In this case, the drive force for a capacitive system scales as $V^2$. The amplitude response of the double-step drive scheme is plotted in Figure 5 b) for a positive as well as a negative deflection angle. Taking note of the x-axis, one can see that the settling time was reduced by almost three orders of magnitude, from ~300 ms to ~320 µs. Consequently, any optical signal passing through the CrossFiber module can be redirected in only 320 µs, a settling time of just 6% above the theoretical settling time limit of $T_0/2$[‡]. There are lower frequency modes visible in the FFT of the ring-down; the transduction is sufficiently weak so as not to interfere with the settling time. The best results were achieved for $F_1 = 0.52\ F_0$ to account for the losses. The CrossFiber LiteSwitch$^{TM}$, a product consisting of an array of 96 photonic switches advertises a typical switching time of 25 ms. It is demonstrated here that this could be reduced considerably without the use of complex and expensive feedback circuits and without increasing the power consumption of the drive circuit.

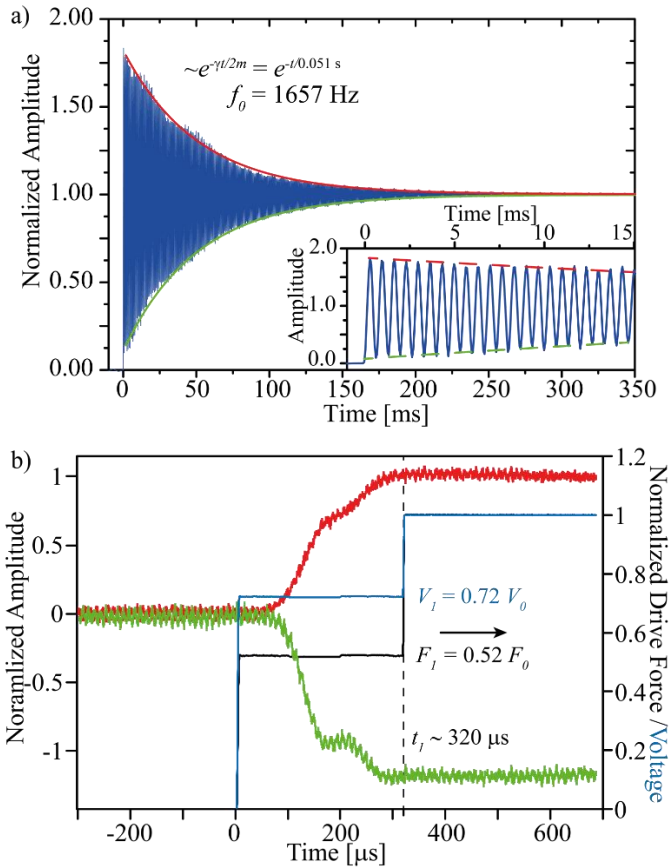

Figure 5. Response of a commercial MEMS optical cross-connect mirror by CrossFiber. a) Single-step drive results in a settling time of almost 300 ms. The envelope of the ring-down is fitted to an exponential, obtaining $2m/\gamma = 51$ ms or $Q = 265$. The inset shows the oscillation over a shorter time interval. The observed beat results from a weaker 550 Hz mode. b) Double-step drive (bi-directional) illustrates the same device settling in just 320 µs, an improvement by three orders of magnitude. The torsional resonant mode was measured using a frequency sweep at 1657 Hz ≈ $1/(2t_1)$ indicating an idealized system could settle in 302 µs, or only 6% faster than experimentally observed.

---

[‡] A frequency sweep revealed the resonance frequency of 1657 Hz and corresponding period of 604 µs.

The second example of an electrostatically driven MEMS device is the Spatial Light Modulator (SLM) seen in the drawing in Figure 2 b). While the previously described electrostatically driven mirrors perform tip-tilt deflections, the SLM is actuated in piston mode. Changing the height of the mirror locally shifts the phase of a reflected wavefront. An array of such independent elements is used in imaging systems such as telescopes and microscopes to correct for wavefront phase errors [55]–[57].

The response of the single and double-step drive of the SLM is illustrated in Figure 6, reprinted from [27]. The SLM consists of a segmented mirror forming an array of square mirrors that are each actuated in piston mode as described in the introduction. Typical displacements are small, roughly half of the wavelength of the reflected wavefront. As a result, the correction to the linear response is small. The response time, however, needs to be as fast as possible. The device is operated in closed-loop mode as the incoming wavefront is corrected dynamically. In principle, this could be used to correct for the dynamic behavior of the mirror; it is, however, significantly simpler to have a well-controlled mirror responding only to the dynamically shifting wavefronts. In this case, the 77.920 kHz segments have a quality factor of 6.3, which is limited by viscous drag. The best results for rapid settling were achieved with $t_1 = 11.1$ µs for the double-step drive, corresponding to an improvement of more than a factor of five. Under certain circumstances, it may be beneficial to operate the device in a vacuum. In this case, the quality factor and hence single-step settling time could rise by two orders of magnitude. The double-step setting time, however, would remain almost unchanged and would even improve from 17 µs to $t_1 = T_0/2 = 6.4$ µs. In this application the mirror speed requirements depend on the frame rate of the wavefront detector. An 11 kHz refresh rate (time constant 91 us) has been demonstrated; where a single-step driven device would be limited by the mechanical settling time, using the double-step drive, the MEMS mechanics described here may allow for significant improvements.

## Electro-Magnetic Actuation: A thousand-fold reduction in settling time of high *Q* micro-mirror devices

Larger, heavier, and slower devices can be efficiently driven by magnetic fields. In contrast to the capacitive drive, this actuation mode can generate larger forces, albeit at the cost of higher actuation power. These large devices may have long ring-down periods due to the significant amount of mechanical energy stored in the moving device, making them ideal for the double-step drive. The example presented here shows how multiple modes of the system can be actuated simultaneously. The resulting ring-down has multiple modes, each with differing frequencies and quality factors. During a step in the applied force, both get actuated and hence a single double-step drive cannot suppress all ringing. However, when sufficiently spaced in frequency, each mode can be tamed individually.

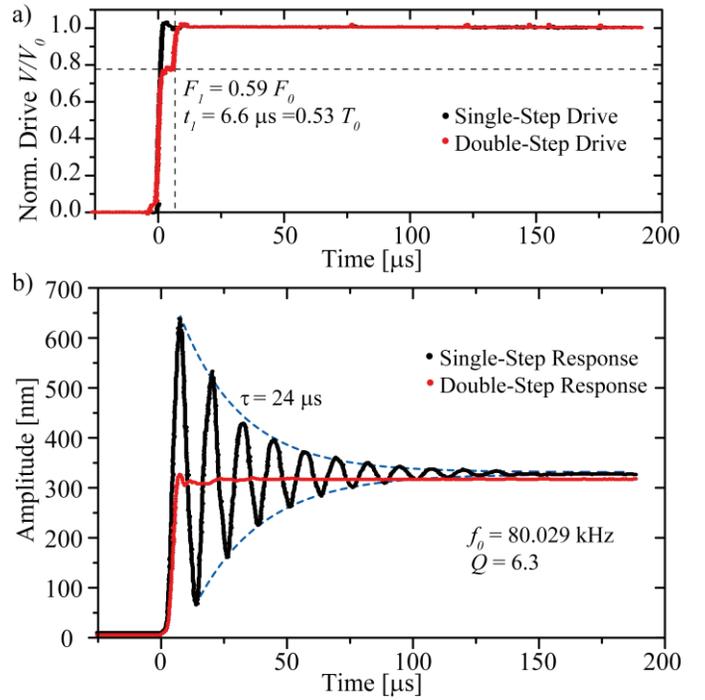

Figure 6. Drive and response of a SLM element. a) Single-step and double-step drive with $t_1 = 6.6$ µs and $F_1 = 0.59\ F_0$. (The analytical solution presented predicts $t_1 = 6.3$ µs and $F_1 = 0.56\ F_0$) The actuation voltage input is normalized to one. b) Single-(blue) and double-step (red) response of SLM element translated by 300 nm. The double-step drive suppresses the ringing below the detection threshold and reduces the settling time from over 100 microseconds to below 17 µs. (Reproduced with permission from SPIE [27]).

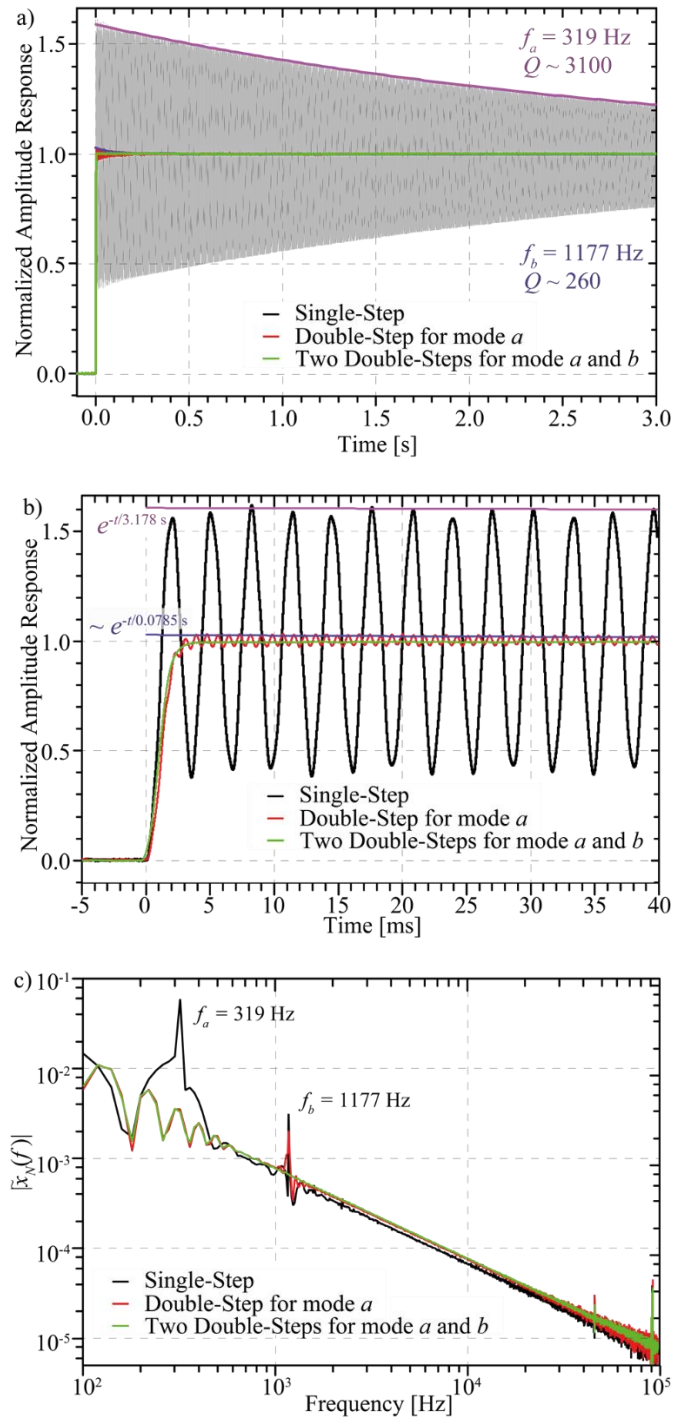

Figure 7. Response of magnetically driven MEMS. a) Single-step, double-step drive and two double-step drive to compensate the first two modes that couple to the applied torque. From the exponential decay, the settling time, $\tau$, and the quality factor, $Q$, can be determined. Using the noise of the amplitude as a scale, it is determined that the single step drive will take 16.8 s to settle to within the noise limit. The two double-step drive settles in under 4 ms, or over 4200 times faster. The minimum time $t_1$ needed to settle based on the double-step drive is $1/(2\times320)$ Hz = 1.56 ms, or 10752 times faster than waiting for the ring-down to complete. b) The initial 40 ms of ring-down after the drive commences. The single-step clearly shows the modulation of the amplitude caused by the higher order mode. The double-step for mode $a$ suppresses the fundamental mode, but significant oscillation of mode $b$ remains. The two double-steps finally results in rapid settling of the MEMS mirror. c) FFT of a) depicts large first and second mode amplitudes for single step drive (black), missing fundamental mode peak but visible harmonic (red) and finally featureless $1/f$ response for the two fold double-step (green).

magnetization $M$ in an electric field $B$, is given by the cross product:

$$\tau_M = M \times B = \perp M \left(\frac{4}{5}\right)^{3/2} \frac{\mu_0 n I}{R} \sin\theta, \quad (11)$$

where the field is generated by Helmholtz coils of radius $R$ through which the current $I$ flows. $\mu_0$ is the permeability of free space, $n$ the number of loops, and $\theta$ the angle between $M$ and $B$. $\perp$ denotes that the torque is perpendicular to both magnetization and field vectors $M$ and $B$. For large amplitudes, the torque is not linear with regard to the actuation angle ($\sin\theta \approx \theta - \frac{\theta^3}{6} + \frac{\theta^5}{120} + \cdots$). As the torque weakens with increasing angle, the current can be increased to compensate. As the magnetization aligns with the magnetic field, the torque vanishes tangentially as expected: No finite current can quasi-statically rotate the mirror beyond $\theta = \pi/2$.

Typical forces can be large compared to capacitive drive techniques, and the required voltages are much lower. The power consumption is high, however, as a current through the coils must be maintained at all times. Furthermore, the MEMS devices become heavy due to the magnets. Further limiting the response time is the high inductance of the drive circuit, slowing the rate at which the torque can be applied. For an inductance-limited circuit, the time constant is $\tau = L/R$ and the impedance scales as $\omega L$; as a result, large dynamic voltages are needed to change the torque at high rates.

The device considered is shown in Figure 2 c). The Sidebar on Finite Element Simulations of resonant modes includes a description of the two modes mentioned previously, which couple to the torque applied to the magnet when a current flows through the coils. The magnetization is a property of a neodymium magnet attached to the backside of the silicon mirror. Adding the magnet to the MEMS increases the mass. Alternatively, coils can be added to the MEMS and the magnetic field can be generated externally [20], [58]. Keeping the coils off the MEMS simplifies the fabrication and removes the input power from the temperature sensitive mechanical elements.

A simple step function drive results in two harmonics being actuated, labeled as mode $a$ and mode $b$. Figure 7 depicts the resulting ring-down, a) and b) in time domain and c) in frequency domain. As each mode is actuated, a double-step drive is required for <u>each</u> mode, which means four parameters need to be determined ($f_a$, $Q_a$, $f_b$, $Q_b$), where high precision measurements are favored over modeling which is rarely sufficiently precise. If only a single double-step drive for the fundamental mode is applied significant ringing in the higher order mode is observed. The two modes are separated by almost a factor of four in frequency space. This makes the device an ideal candidate for two fold double-step actuation. The high frequency mode can settle within each level of the double-step drive used to suppress the ring down in mode $a$. Essentially, two point-to-point transitions are completed, each requiring a double-step drive. Given the large difference in period, $t_{1a} \gg t_{1b}$, the double-step for mode $b$ is implemented sufficiently quickly so that mode $a$ is oblivious to it. For this to

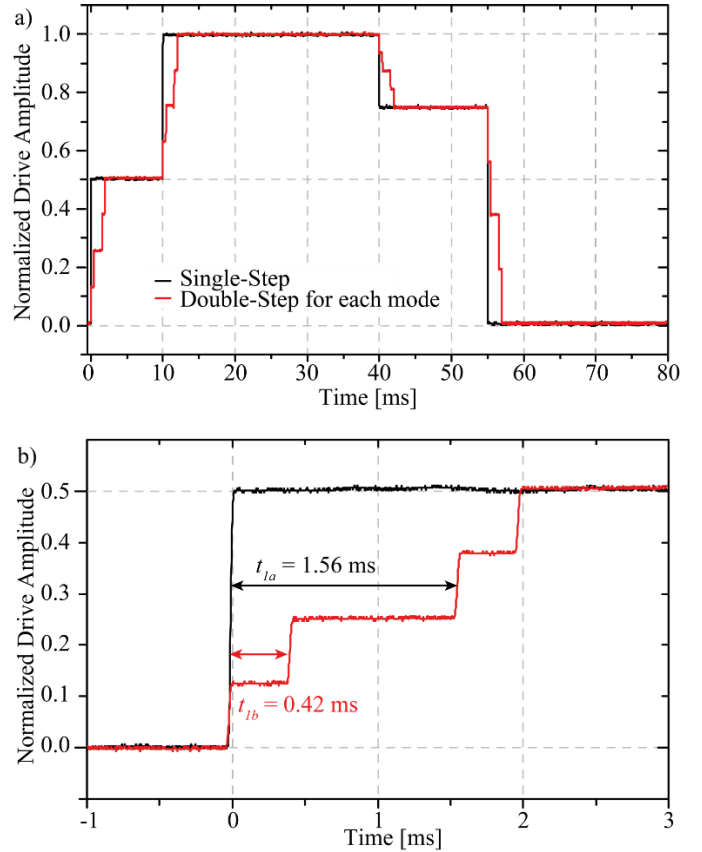

Figure 8. Magnetic drive transduction. Two modes couple into the step function. a) Normalized drive of single-step and two double-step actuation schemes, for switching between five amplitudes. b) First transition from 0 to 50 % of the maximum amplitude. Each transition of the double-step is split into two to account for both modes.

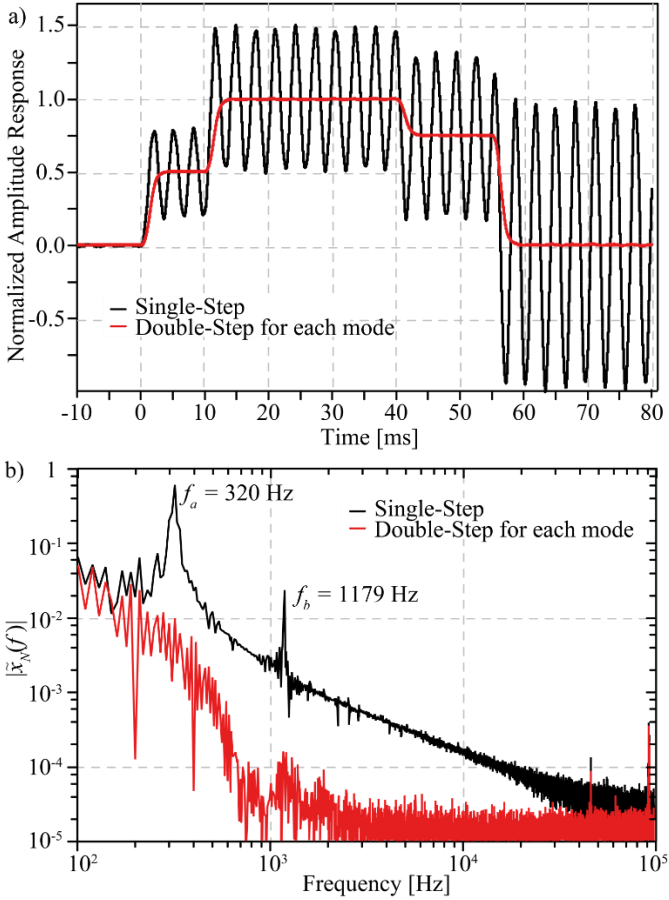

Figure 9. Amplitude response of magnetic drive. Point-to-point transitions between five amplitudes. a) Normalized response of the driving force depicted in Figure 8. Single-step forcing between levels results in significant long term ringing. A double-step for each mode results in a smooth, oscillation-free response. b) FFT of a) depicts large first and second mode amplitudes for the single-step drive. The double-step for each mode completely eliminates the frequency peaks at 320 Hz and 1179 Hz.

Figure 7 shows the settling of both modes on the order of 4 ms, considerably longer than the theoretical 1.5 ms defined through the fundamental mode period of 3.1 ms. The reason for this is the increased duration resulting from the two double-step drive (adding 0.42 ms to the step duration) and the time constant of the inductive circuit. Acting as a low pass filter, the magnetic coils limit the rate at which the force can be modulated. The positioning time, defined as the duration until the ringing drops below the noise amplitude, is found to be 4200 times faster for the two double-step drives than for the single-step drive. The efficient suppression is demonstrated by the FFT plots in Figure 7 c). Shown are strong peaks at 319 Hz and 1177 Hz for the single-step drive, a single 1177 Hz peak for the double-step drive, and no peaks for the two double-step drives. Although at frequencies approaching $10^5$ Hz, additional features in the FFT spectrum are visible, these are too weak to cause any measurable displacements. With quality factors of 3100 and 260, the dissipation is negligible and, as expected, $F_1 = F_0/2$ for both modes. The feedforward filters, described in detail in the Drive Filters Sidebar, is well suited to discover how to treat multimodal responses. Each mode requires its own filter (and transfer function). Simply multiplying the two filters (the formal way of applying both filters) produces the desired minimum time settling. Each filter has two terms, the multiplication of which results in four terms corresponding to the four transitions shown in Figure 8 b). The example shown here is a special case where only two modes need to be considered.

Figure 8 and Figure 9 depict the forcing and position responses for four point-to-point transitions, two increasing and two decreasing in amplitude. Both single- and two double-step transitions are considered. The effectiveness of the drive is truly impressive. In applying the two double-step forces, all four transitions are accomplished within 60 ms (with an experimental minimum limit of 4×4 ms = 16 ms), whereas the single-step response continues to ring uninterrupted. Considering a mechanical settling time of over 16 s, controllably addressing all four set-points would take over a minute. Again the FFT of the response is plotted to demonstrate the work using this approach the higher order frequency must be at least double the fundamental mode frequency. This condition results from the requirement for the higher order mode to settle during the first step of the actuation of the fundamental mode. When using the feedforward filters to determine the input command there is no restriction on the frequency spacing of the modes. As discussed in the drive filter sidebar, the input command can be generated for a system with an arbitrary number of modes of any frequency sequence. The final settling time, $t_{TOT}$, will be the sum of the settling times of the individual modes ($t = \sum t_{1i}$).

effective suppression of all resonant modes. The high-frequency suppression in the FFT is discussed in detail below, where the effect of smoothed drive forces resulting in curbed higher order modes is illustrated.

While mechanical modes above 1177 Hz may exist for this system, they are not actuated as the coils act like a low-pass filter and do not allow for high frequency actuation. This can be exploited technologically because the response time is limited by the $t_1$ of the fundamental mode and higher order modes can be suppressed by simply integrating a low-pass filter into the drive electronics. If the higher order modes are too close in frequency for a low-pass filter to not influence the fundamental mode, then the twofold double-step drive can be implemented. The system must be engineered to exclude higher order modes too close in frequency to the fundamental mode. A simulation of a double-step torque for suppressing multiple resonant modes is presented by in reference [59] for the intended application of reorienting spacecraft.

**Thermoelectric Forcing: Generating strain gradients to shape and orient mirrors**

There are two types of thermal drive schemes. In one, a temperature gradient is established where the resulting differential change in length of the two elements results in a deformation [60]. In the second, a bimorph, typically a stack of silicon and gold, is heated. The resulting difference in thermal expansion results in a strain gradient across the structure, which then deforms mechanically. The bimorph can be characterized by its radius of curvature which can both shape the surface of a mirror or be used to control its angle or height [15]. Considering the initial curvature, $\kappa_0$, any change in temperature, $\Delta T$, will result in a change of curvature given by [61]

$$\kappa = \frac{1}{r} = \kappa_0 + \frac{6t\Delta\alpha\Delta T}{4(t_{Au}^2+t_{Si}^2)+6t_{Au}t_{Si}+\frac{E_{Au}t_{Au}^3}{E_{Si}t_{Si}}+\frac{E_{Si}t_{Si}^3}{E_{Au}t_{Au}}}, \quad (12)$$

where $r$ is the radius of curvature, $\Delta\alpha$ the difference between the thermal expansion coefficients of silicon and gold, $t_i$ the thicknesses, and $E_i$ the Young's moduli of both materials. The angle of the mirror is proportional to the curvature $\kappa$ as $\theta = L\kappa$, with $L$ the length of the bimorph. Not only dependent on the mechanics of the device, the dynamics are also governed by the rate at which the bimorphs can be heated and cooled. Assuming the mirror cools to air and thereby forms a heat sink [62], the thermal relaxation time $\tau_{th}$ of the device is given by

$$\tau_{th} = \frac{L^2 \rho C_p}{\pi^2 \kappa_{th}}, \quad (13)$$

where $L$, $\rho$, $C_p$ and $\kappa_{th}$ are the effective length, density, heat capacity and thermal conductivity of the bimorph respectively. The $L^2$ scaling means that as the device size shrinks, the thermal timescale becomes very fast. Consequently, NEMS devices can be thermally driven at frequencies exceeding 200 MHz [63]. Using effective parameters to account for the gold and silicon materials used in the bimorph, equation 12 results in a thermal time constant of $\tau_{th}$ = 1.1 ms, in agreement with the thermal timescale measurements illustrated in the Sidebar on Thermal Relaxation and Piezoresistivity. Typically, a device is heated thermoelectrically while the cooling is passive. For a conductive bimorph, this means that heat can be generated throughout the structure, where cooling is both through the surrounding air and into the base. Decreasing the thermal conductivity lowers the power consumption, but also increases the thermal relaxation time. For the double-step drive to work effectively, the device must limited by the mechanical, and not thermal, timescale.

The change in temperature results from an electrothermal drive, where Joule heating in the bimorphs equilibrates through thermal dissipation into the silicon substrate and the surrounding air. For small temperature changes, $\Delta T$ is proportional to the applied power, a function of the square of the drive current (or voltage depending on how the device is biased)

$$P = I^2 R(T) = \frac{V^2}{R(T)}. \quad (14)$$

It must be noted that the resistance, $R(T)$, itself will rise with increasing temperature. For a current-biased drive, this results in additional heating and can lead to thermal runaway and catastrophic failure. For a voltage-limited drive, there is an initial power peak, which then drops as the resistance rises with temperature. For this reason, a voltage-biased system will typically respond faster than a current-biased one.

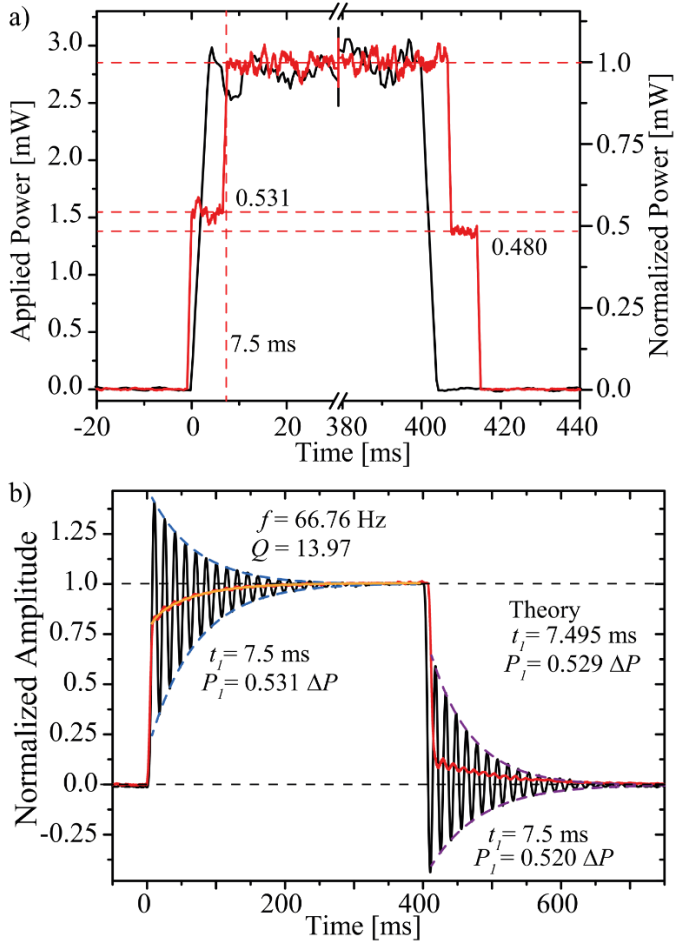

Figure 10. Electrothermal drive of large angle mirror a) Drive and b) response for single- and double-step forcing respectively. The exponential decay and fit to the oscillation reveals a resonance at 66.76 Hz and corresponding quality factor of 13.97. The derived theoretical values for $t_1 = 7.495$ ms and $P_1 = 0.529$ $\Delta P$ are within 0.1 % and 2 % of the best experimental values found respectively. A thermal drift is characterized by two relaxation time constants $\tau_1 = 10$ ms and $\tau_2 = 64$ ms illustrated by fit (the orange trance) to the oscillation free response.

The thermally driven MEMS device shown in Figure 2 c) consists of two gold-silicon bimorph actuators with a large mirror attached to their ends. Intrinsic strain from the fabrication process results in an elevated structure. Heating the device flattens the bimorphs and thus changes the angle of the mirror, similar to previously reported optical MEMS microscanners [21], [64]. The single-axis device exhibits a large angular range, positioning a mirror with a ~1 mm² surface area. The plot in the Thermal Relaxation and Piezoresistivity Sidebar show that by applying a voltage bias while measuring the current, two thermal relaxation times can be observed. The first one, related to the thermal relaxation time for the device itself, is determined to be 0.95 ms, and a second, longer relaxation time, probably determined by the substrate-bimorph-mirror system, is measured to be 39.5 ms. Furthermore, a mechanical ringing is observed: the thermal pulse causes the mirror to resonate. The modulation of the strain is observed in the oscillating current resulting from the piezoresistive properties of the device (details are given in the sidebar). In certain cases, this modulation works as a heat engine and can drive a MEMS device [65].

This ringing is also visible in the reflected light of a laser onto a position sensitive detector (PSD) as shown in Figure 10. The single step response is again used to determine the electro-mechanical coupling as well as the parameters $f_0$ and $Q$. The oscillations are accompanied by an additional slower relaxation, or drift, as is observed in the resistance. While the ringing can be suppressed by the double-step drive, the drift (with a mechanically measure timescale of 64 ms, see orange trace in Figure 10 b)) must be mitigated using other methods. One possibility is an intentional thermal drive overshoot followed by a slow relaxation to the desired final drive amplitude. Such drive sculpting is no longer described by the linear theory presented here. While characterizing the device can lead to improved open-loop forcing, high accuracy in such devices is only achieved in a closed-loop mode. It should be noted that thermally overdriving the device produces an increase in the ringing. Consequently, even in a closed-loop setup, the double-step drive can minimize overshoot and simplify the feedback, as the resonant response has already been suppressed and only the much slower transient corrections are required. A plot of the drive power (Figure 10 a)) shows that applying just over half power for half the period ($t_1 = \frac{1}{2}T_0 = 7.5$ ms) effectively suppresses the resonant response. The much slower thermal relaxation time can be dealt with by intentionally overshooting the applied power and then smoothly ramping it back for the desired final position. Even in this highly non-linear system, the theory predicts the correct values for $t_1$ and $P_1 \sim F_1$ within 0.1 % and 2 % respectively. It can be expected that for higher drive amplitudes, the nonlinearities will become increasingly significant, resulting in a deviation from the simple theory given here. For ringing to occur at all, the thermal timescale must be significantly shorter than the period of the mechanical resonance being actuated, given a thermal timescale of just over one millisecond, this is certainly the case for the example presented here.

While thermal forces can be very large and usefully applied to MEMS devices, the power requirements are higher than for many other drive methods. While thermally driven MEMS are more efficient when operated in vacuum; the thermal timescales will increase correspondingly. Another drawback of thermal actuation is the long-term stability of the system; elevated temperatures can cause material degradation and fatigue. As an example, silicon and gold can form a silicide at only 350 C, a temperature easily reached in MEMS systems, resulting in device failure.

# Large force actuation and resonant drive schemes

### Overdrive: High-speed opportunities and limitations when $F \gg kx$

Considering the maximal force, $F_m$, which can be applied to a given system (typically the result of the maximum voltage or current available), one can imagine a response time faster than that determined by the restoring force of the spring. Here two situations are considered. For overdrive $a$ both an accelerating and decelerating maximal force ($+F_m$ and $-F_m$ respectively) can be applied to the mirror. $F_m$ can greatly exceed $F_0$ and is required to be greater or equal to $F_0/2$. As a second example, termed overdrive $b$, one may consider a system which only allows for an externally applied acceleration, and the mechanical spring provides the required restoring force to bring the device to rest. An example of such a system would be the thermal drive, the heating power can be turned on and off, but no active cooling is applied. Again the system is described by the maximal externally applied force $+F_m$.

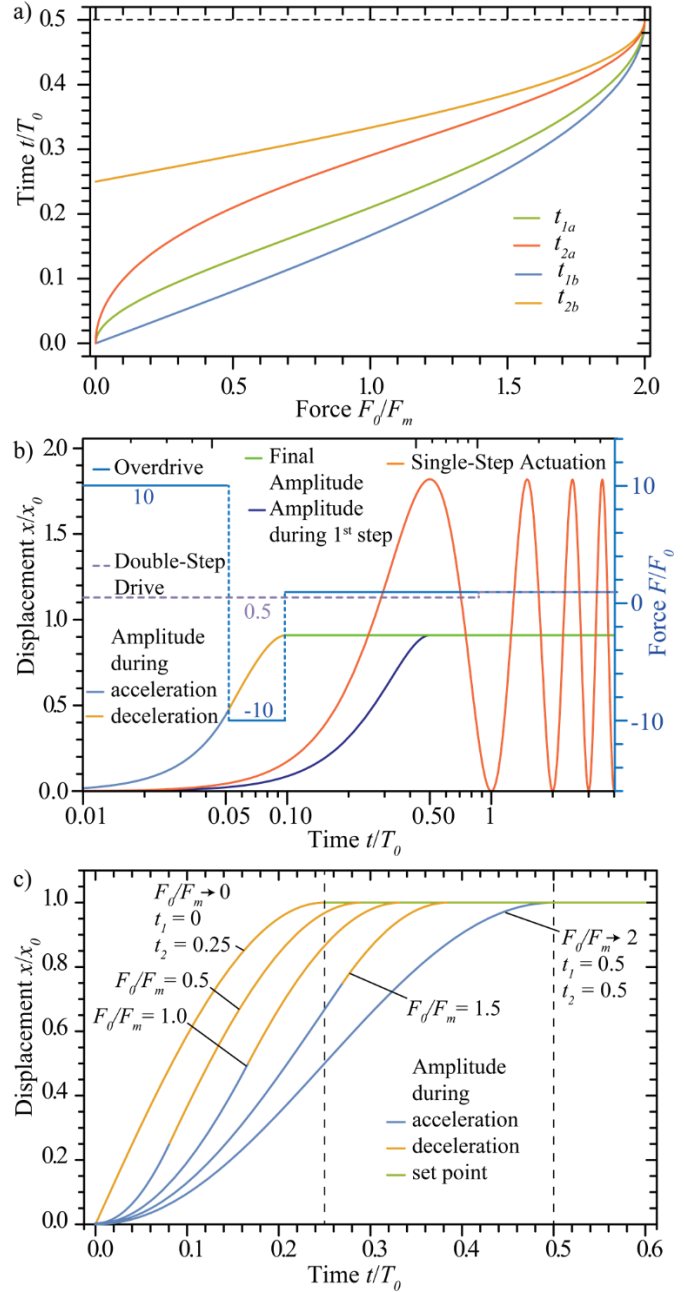

Figure 11. Overdrive mode. a) Duration of acceleration and deceleration as a function of $F_0$, given a final resting position of $x_0 = F_0/k$. Overdrive $a$ both $t_{1a}$ and $t_{2a}$ vanish as the applied force diverges. For Overdrive $b$ a minimum settling time of $T_0/4$ is obtained for a divergent drive force. b) Comparison of single-, double-step, and Overdrive $a$. Assuming a maximum applied force of $10 \times F_0$, the settling time can be reduced to $0.0987 \times T_0$, or almost five times faster than the double-step settling time. c) Illustration of Overdrive $b$. As the accelerating force increases its duration shortens to allow for the mechanical spring to bring the device back to rest at the new position. Without an external deceleration force a minimal settling time of $T_0/4$ can be reached. For both Overdrive $a$ and $b$ the double-step drive is recovered as $F_m \to F_0/2$.

Consider overdrive $a$: In the dissipation-free limit, the acceleration of a MEMS device becomes (setting $\gamma = 0$ in equation (3))

$$\ddot{x} = \frac{F_m}{m} - \frac{k}{m}x. \tag{15}$$

The general solution of the inhomogeneous differential equation can be expressed as

$$x = \frac{F_m}{k} + c_1 \cos\sqrt{\frac{k}{m}}t + c_2 \sin\sqrt{\frac{k}{m}}t. \tag{16}$$

Remembering that the aim is to accelerate and then decelerate the mirror with a maximal force of $\pm F_m$, one must determine how long to accelerate and then decelerate the device. For a system initially at rest at the origin, the position of the resonator during the acceleration phase becomes

$$x_{aa}(t) = \frac{F_m}{k}(1 - \cos\sqrt{\frac{k}{m}}t). \tag{17}$$

During the deceleration the position becomes

$$x_{da}(t) = \frac{F_m}{k} + c_{1a} \cos\sqrt{\frac{k}{m}}t + c_{2a} \sin\sqrt{\frac{k}{m}}t, \tag{18}$$

where $c_{1a}$ and $c_{2a}$ are constants determined by the boundary conditions. The time $t_{1a}$ is defined to be the end of the acceleration and the time $t_{2a}$ to be the time at which the deceleration is completed and the device is positioned at the target location and at rest. This imposes the following four boundary conditions needed to be solved for the four unknowns

$$x_a(t_{1a}) = x_d(t_{1a}), \tag{19}$$

$$\frac{dx_a(t_{1a})}{dt} = \frac{dx_d(t_{1a})}{dt}, \tag{20}$$

$$x_d(t_{2a}) = x_0, \tag{21}$$

$$\frac{dx_d(t_{2a})}{dt} = 0, \tag{22}$$

from which all the remaining parameters can be calculated

$$t_{1a} = \frac{T_0}{2\pi}\cos^{-1}\left(\frac{4F_m^2 - 2F_mF_0 - F_0^2}{4F_m^2}\right), \tag{23}$$

$$t_{2a} = \frac{T_0}{2\pi}\cos^{-1}\left(\frac{2F_m^2 - 2F_mF_0 - F_0^2}{2F_m(F_m+F_0)}\right), \tag{24}$$

$$c_{1a} = \frac{2F_m^2 - 2F_mF_0 - F_0^2}{2F_mk}, \tag{25}$$

$$c_{2a} = \frac{\sqrt{F_0(2F_m-F_0)(2F_m+F_0)(4F_m+F_0)}}{2F_mk}. \tag{26}$$

Figure 11 a) shows the calculated values for $t_{1a}$ and $t_{2a}$ as a function of $F_0/F_m$, as well as the expected amplitude response for a forcing of ten times the final equilibrium force (see Figure 11 b)). It should be noted that in the limit $F_m \to F_0/2$ the overdrive solution converges to the double-step solution where $t_{1a} = t_{2a} = T_0/2$, and no deceleration force is applied (the duration during which the applied force is $-F_m$ is $t_{1a}-t_{2a}$ and vanishes in the $F_m \to F_0/2$ limit).

From a practical point of view, the applicability of the overdrive method is limited. Many MEMS devices cannot withstand maximal forces of an order of magnitude higher than the required steady-state forces they will experience. As large angles are desired, technologically $F_m$ is often the same order as the full range of $F_0$. Also, short, high-amplitude pulses result in driving forces acting at very high frequencies. Consequently, higher-order modes of the structure may be actuated; so even if no ringing at the fundamental frequency is observed, there may be significant higher-order modes actuated with long settling times. Nevertheless, there are some specific applications where an overdrive approach can be implemented. For example, short, narrow, but relatively thick torsion springs have a reasonably low torsional spring constant. The fundamental torsion mode is well separated in frequency space from the next order torsion mode that can couple into an applied torque (see Sidebar on Finite Element Simulation of is expected to be $T_0/14.2$, quite close, given that the theory is based on a point-mass with a massless spring system while the finite element simulation takes the entire three-dimensional device into consideration. The agreement between the theory and this simulation is based on two important factors. First, the torque applied is ideal, meaning there are no translational forces that can actuate modes orthogonal to the torsional mode. This is experimentally difficult to achieve, as any misalignment of the drive force will increase coupling to modes orthogonal to the desired torsion mode. Second (and related to the first point), the torsion mode is unique in that higher-order torsion modes that could be actuated by a pure torque are at very high frequencies. For the geometry considered here, the next order torsion mode is at 452 kHz, over two orders of magnitude above the fundamental mode. Furthermore, only the springs participate in the motion, while the mirror plate at the center remains still; hence even if actuated, this is not a mode that can cause an unwanted beam deflection. The first higher-order mirror torsion mode was found to be at 1.358 MHz, far too high to be actuated by the slow drive circuit considered here.

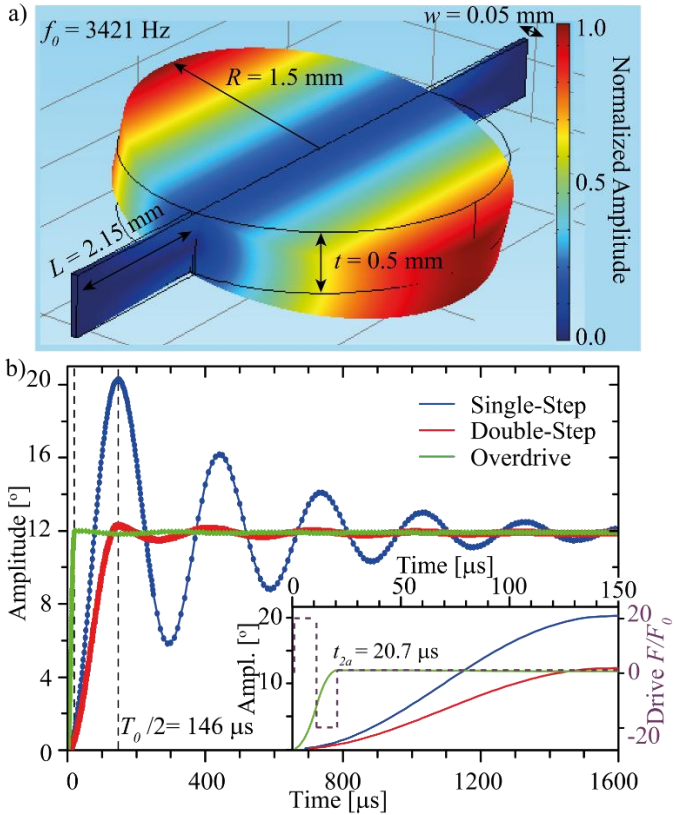

Figure 12. Simulation of a single-axis mirror. a) The fundamental torsional mode appears at 3421 Hz. b) Single-(blue), double-step (red), and Overdrive (green) actuation. The Overdrive results in a settling time seven times faster compared to the double-step actuation and is completed after only $t_{2a} = 20.5$ μs $= T_0/14.3$ for $F_m = 20\,F_0$.

Resonant Modes). Given that linear modes in the low dissipation limit do not couple, if the torque is applied effectively, only the desired resonance can be actuated.

The overdrive $a$ method is demonstrated for the single-axis mirror using finite element simulations, illustrated in Figure 12. In a), the fundamental torsion mode is reproduced, with colors indicating relative amplitudes at a resonant frequency of 3421 Hz; b) illustrates the amplitude response in the time domain for the single, two, and three step drive schemes. As predicted, the double-step settles after $t_{1a} = 148$ μs $\approx T_0/2$. Some ringing is still observed as the simulation is sensitive to the discretization of time. Applying the overdrive with the torque $\tau_m = 20\,\tau_0$ results in a settling time of $t_{2a} = 20.5$ μs $= T_0/14.3$. In this case, the ringing is practically nonexistent. Considering the predictions given by (23) and the plots in Figure 11 a), the settling time

In the magnetic drive example illustrated above, the magnet on the device introduces an asymmetry. The rotational axis is not the center of mass and, when applying a torque to the magnet there are components acting on multiple modes. This is illustrated by the peaks of the FFTs plotted in Figure 9 b), where the corresponding mode shapes are illustrated in the Sidebar. Even if the modes could be sufficiently isolated, there is still the difficulty of generating the required high torques. Considering the drive modality, the high currents needed to generate $\tau_m$ cannot be applied continuously, as the coils would overheat and melt. It may be possible to pulse the coils for $t_{1a,2a}$ of order $T_0/20$ without catastrophic meltdown. Of course, if only very small, yet high speed amplitude corrections are needed, the requirements on $F_m$ are eased, and the overdrive actuation scheme becomes more widely applicable.

Although the derived analytical model presented here applies only to the vanishing dissipation limit, this is not a particularly severe constraint, since MEMS device can often be operated in this limit. It should be noted that the dissipation is proportional to the velocity. Hence, even though the device moves quickly before settling, the correspondingly high speeds result in an energy loss equivalent to that of the double-step drive method.

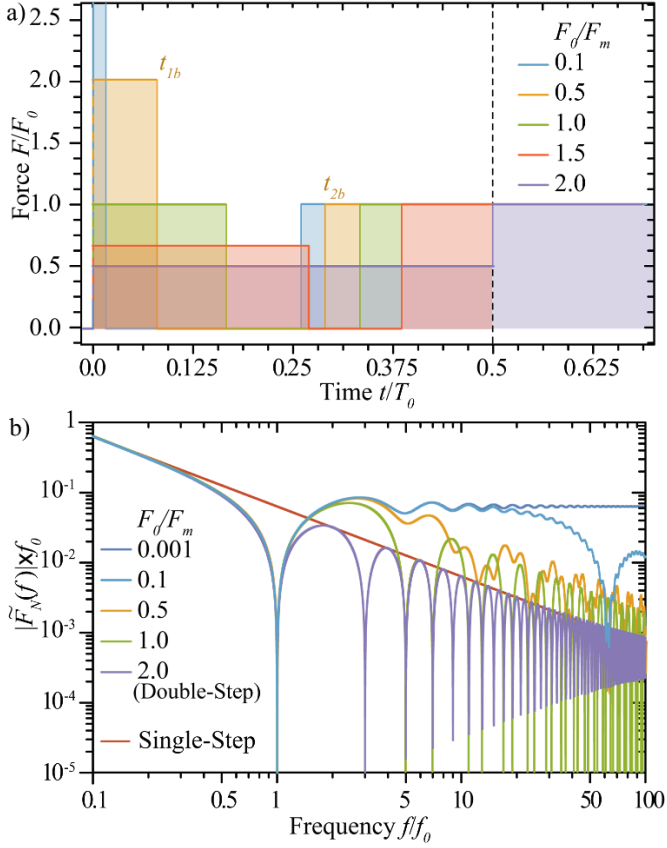

Figure 13. Actuation traces and their Fourier transforms of Overdrive b. a) Five Overdrive forcing examples ranging from $F_m = 10\ F_0$ to $F_m = F_0/2$. b) Fourier transform of Overdrive b drive force. As is the case for Overdrive a, the resonant mode can be suppressed from any drive force. The shorter time pulses for increased $F_m$ result in higher actuation forces acting at higher frequencies.

Where the double-step drive provides a smooth point-to-point transition, it is not optimized for time. The overdrive a input shaping presents the minimal time solution given a maximal bi-directional force. One may also consider the minimal time solution for single sided forcing, overdrive b. In such cases, the aim must be to accelerate the system to the largest possible velocity for which the spring of the MEMS can still provide sufficient restoring force to bring the device to a halt by the time it reaches the desired amplitude. Again, there will be a period of acceleration during which the force $+F_m$ is applied, followed by a deceleration period during which the restoring force $-kx(t)$ acts. The actuation profile of such a drive mechanism is illustrated in Figure 13 a). The resulting expression for the amplitude response of the acceleration phase is identical as for the overdrive a case (see equation (17)).

$$x_{ab}(t) = \frac{F_m}{k}\left(1 - \cos\sqrt{\frac{k}{m}}t\right). \qquad (27)$$

During the deceleration, the applied force is set to zero and hence

$$x_{db}(t) = c_{1b}\cos\sqrt{\frac{k}{m}}t + c_{2b}\sin\sqrt{\frac{k}{m}}t, \qquad (28)$$

The boundary conditions are as before, where $t_{1b}$ is the duration of the acceleration force and $t_{2b}$ is the settling time at which force $F_0$ is applied.

$$x_a(t_{1b}) = x_d(t_{1b}), \qquad (29)$$

$$\frac{dx_a(t_{1b})}{dt} = \frac{dx_d(t_{1b})}{dt}, \qquad (30)$$

$$x_d(t_{2b}) = x_0, \qquad (31)$$

$$\frac{dx_d(t_{2b})}{dt} = 0, \qquad (32)$$

from which all the remaining parameters can be calculated

$$t_{1b} = \frac{T_0}{2\pi}\left(\tan^{-1}\left(\frac{2-\left(\frac{F_0}{F_m}\right)^2}{2\frac{F_0}{F_m}\sqrt{1-\left(\frac{F_0}{2F_m}\right)^2}}\right) + \frac{\pi}{2}\right), \qquad (33)$$

$$t_{2b} = \frac{T_0}{2\pi}\left(\tan^{-1}\left(\frac{\frac{F_0}{F_m}}{2\sqrt{1-\left(\frac{F_0}{2F_m}\right)^2}}\right) + \frac{\pi}{2}\right), \qquad (34)$$

$$c_{1b} = \sin\left(2\pi\frac{t_{2b}}{T_0}\right), \qquad (35)$$

$$c_{2b} = \cos\left(2\pi \frac{t_{2b}}{T_0}\right). \tag{36}$$

Figure 11 a) shows the calculated values for $t_{1b}$ and $t_{2b}$ as a function of $F_0/F_m$. As $F_m$ diverges the duration of the acceleration vanishes. The velocity reached essentially instantaneously will subsequently decreased due to the breaking force of the spring until the desired amplitude $x_0$ is reached at time $t_{2b} = T_0/4$, the minimal settling time of a resonator with unidirectional forcing. For $F_m = F_0$ the settling time reaches $T_0/3$. Just as before, in the $F_m \rightarrow F_0/2$ limit, the overdrive solution converges to the double-step solution where $t_{1a} = t_{2a} = T_0/2$. The corresponding amplitude curves are plotted in Figure 11 c).

Where here only the analytical solution is presented, studies on capacitive comb drive MEMS [31] compare the responses of single-step, overdrive $b^§$ command sculpting (feedforward) and closed-loop (feedback). Where the closed-loop improves the single step rise time from 190 µs to 170 µs, it falls short of the overdrive (pre-shaped) response time of 100 µs. The actual settling time improvement of the feed forward and closed-loop drive schemes are by a factor of roughly three times faster than standard open loop drive. The observed timescales are in agreement with the overdrive $b$ scheme using $F_m \approx 20\, F_0$. The closed-loop response time in this case was limited by the sampling rate.

The functions obtained for of the overdrive actuation schemes can be used to determine the feedforward input filters used in command theory. The results are discussed in the sidebar on Feedforward and Feedback Drive Fitters, and are used to corroborate the discussion on sensitivity given below.

**Resonant drive schemes**

So far, only various sets of step functions were presented for driving the resonator. There are other drive modalities that may be considered. Instead of generating a large overshoot which rings down to a new equilibrium as described above, a mode driven harmonically (possibly on resonance) will steadily gain amplitude until the dissipation prevents a further increase. A sinusoidal drive is proposed in [66] for vibration free point-to-point transition. One can consider a drive force that results in a local maximum being reached at the desired amplitude. At the apex, the resonator is at rest; hence, if at that moment the forcing is switched from harmonic to a constant, the restoring force is balanced and the position can be maintained. The drive force and the resulting displacements for this scenario are plotted in Figure 14 a) and b) respectively. In one case, the steady state is reached at the first maximum (or second root of the velocity) and in another case the desired amplitude is reached only at the third maximum (or $6^{th}$ root of the velocity). The higher the forcing, the more rapidly the amplitude is generated. Just as for the double-step drive scheme, the dissipation and resonance frequency will define the required experimental parameters. In the example presented here, given a dissipation of $\gamma \approx$ 0.4, the first maximum is reached at 0.78 $T_0$, not much longer than the settling time of the double-step drive. However, the required forcing is approximately $3F_0$. If one is more patient, a resonant drive of $0.9F_0$ will reach the steady state amplitude of $F_0/k$ by $t = 2.78\ T_0$.

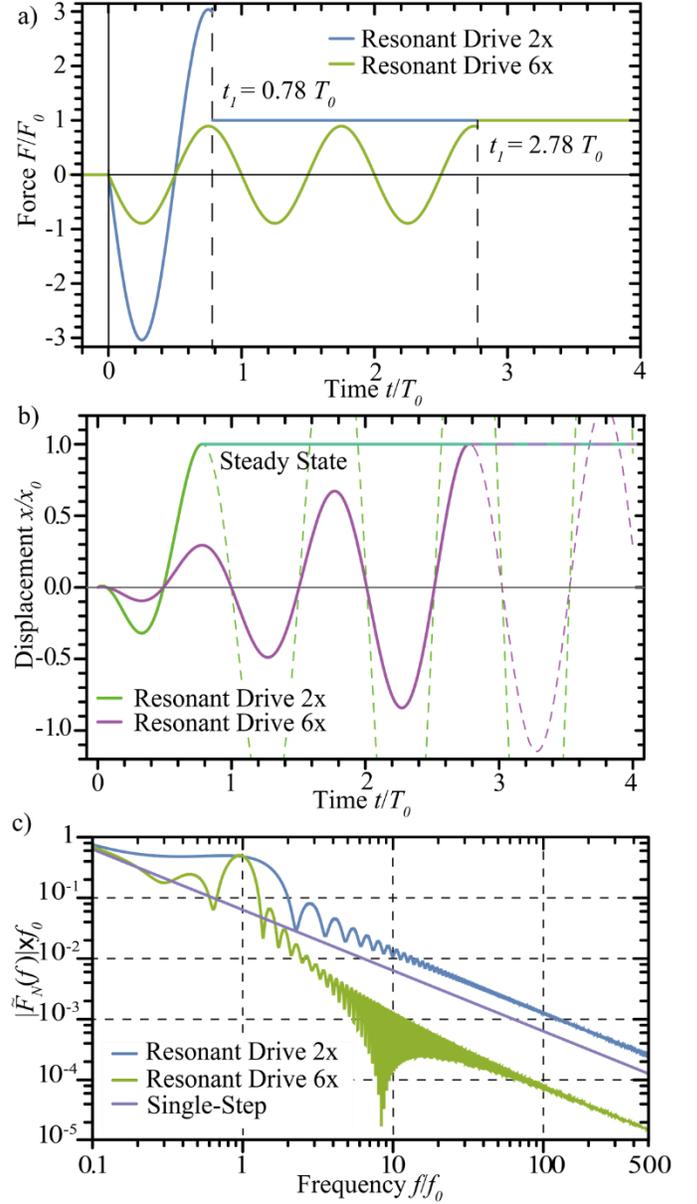

Figure 14. Double-step resonant drive actuation. a) Resonant ring-up followed by constant forcing for two resonant drive durations. $n$x refers to the number of zero crossings the resonator performs before it is held in place. b) Displacement response for the two resonant drive actuation forces depicted in a). c) Fourier transforms of the drive force. Longer resonant drive results in narrowing features in the frequency domain. Each drive is completed by a step function, consequently all frequencies are addressed in addition to the brief resonant drive. For 6x resonant driving the ring-up amplitude is close to the final amplitude where the constant forcing is applied. Consequently, the amplitude of the step is small and the higher frequencies are suppressed compared to the single step or 2x resonant drive.

---

§ Overdrive $b$ is well tailored to capacitive drive schemes: Large voltages can be set very quickly but the resulting forces act only in one direction.

At first glance, one may believe that the resonant drive will selectively actuate only a specific frequency and is hence desirable over the indiscriminate double-step drive that can actuate other modes. However, given that the resonant drive is always concluded by a step to the final drive force, there will, by necessity, be significant contributions to the forcing at all frequencies. As depicted in Figure 14 c), the resulting FFT of the actuation forces considered shows drive amplitudes in frequency space equivalent to that of the single-step function. One special exception would be if the amplitude driven on resonance with the drive force $F_0$ reaches an apex amplitude of $F_0/k$. In this case, the step function can become arbitrarily small and the higher order frequency contributions are weaker. Although such a system may be constructed, it is unlikely to be of any practical value as only specific parameters over a narrow range can fulfill this condition.

## Stability of the Double-Step and Overdrive Actuation Schemes

For the double-step input sculpting to work effectively precise knowledge of the device parameters must be known. For the linear systems described in this article there are only two parameters, the resonance frequency and the dissipation. It is assumed that the electromechanical transfer function is known precisely, i.e. any position can be reached with arbitrary precision. Theoretically these parameters can be calculated analytically or numerically, in practice, as is the case for all examples discussed here, the resonances $f_i$ and quality factors $Q_i$ are determined experimentally. Two approaches can be used: either a frequency sweep fitted to the Lorentzian curve, or a step force can be applied, the ring-down recorded and fitted to equation (5).

In this section the sensitivity of the double-step drive to de-tuning, or errors, on the system parameters is considered. Such errors can arise due to imperfect characterization or as a result of changes in the MEMS devices. Where resonance frequencies can easily be measured to 1 ppm, thermal fluctuation and aging effects will cause the resonance frequency to drift. For example, a device may be expected to function over a 100ºC temperature range over which a silicon device is expected to exhibit a frequency shift of order 1% [67], where need careful design and material choice can significantly reduce the thermal sensitivity of MEMS devices [68]. More significant changes due to environmental effects, such as changes in humidity, are typically avoided using hermetic packaging.

The stability of the double-step drive is illustrated by introducing errors in the time, $t_1$, force, $F_1$, and the precision of the input sculpting, where the error is introduced through smoothing the heavy side function. Errors in $t_1$ not only encompass experimental limits but are also equivalent to errors or drift in the resonance frequency where errors in $F_1$ may be attributed to changes in the quality factor or changes in the electromechanical coupling. Smoothing of the input command can result from imperfect drive electronics (finite ramp rate of a voltage source) as well as limiting timescales due to

## Effects of errors in amplitude and time

*Figure 15* a) depicts the response error, measured by the overshoot normalized to the desired set point. The detuning variables $t_{ERR}$ and $F_{ERR}$ are introduced which describe the deviation from the optimal values, $t_{1c}$ and $F_{1c}$, where no ringing or overshoot occurs. Hence, the double – step drive is characterized by $t_1 = t_{1c} + t_{ERR}$ and $F_1 = F_{1c} + F_{ERR}$. The sensitivity to $t_{ERR}$ and $F_{ERR}$ is studied using the same finite element simulation as was used to demonstrate the overdrive *a* minimum settling time actuation discussed in the previous section. *Figure 15* b) and c) show the amplitude response with respect to de-tuning $t_1$ and $F_1$ respectively. The overshoot in a) is determined from the maximum value of these simulations. The trend shows that for maximal detuning ($t_{ERR} = \pm T_0/2$ and $F_{ERR} = \pm F_0/2$) the overshoot is the same as the set point amplitude, and per definition the overshoot vanishes for $t_{ERR} = 0$ and $F_{ERR} = 0$. The red and blue solid traces in *Figure 15* a) result from plotting the amplitude of the Fourier transformation of the forcing term, normalized to the expected detuning at the extremes. Once can show that the overshoot follows

$$\frac{x_{ERR}}{x_o}(t_{ERR}) = \left|\sin \pi \frac{t_{ERR}}{T_0}\right|, \quad (37)$$

$$\frac{x_{ERR}}{x_o}(F_{ERR}) = \left|2\frac{F_{ERR}}{F_0}\right|. \quad (38)$$

This means that a 1% error in time (or frequency) results in a 3.1% overshoot, correspondingly a 1% error in forcing results in a 2% overshoot. The solid diamonds in *Figure 15* a) depict overshoot data for the capacitive cross-connect mirrors with regards to errors in timing. The trend matches the simulated and theoretical values, the offset can be explained by detuning of other parameters not associated with time.

The analogous finite element simulations and the Fourier transformations of the overdrive *a* and *b* actuation function reveal, not surprisingly, a much stronger sensitivity to timing errors. It can be shown that in the high forcing limit ($F_m \gg F_0$) errors in $t_{1a,b}$ result in an overshoot of the form $\frac{x_{ERR}}{x_o}(t_{ERR}) = 4\pi \frac{F_m}{F_0}\left|\frac{t_{ERR}}{T_0}\right|$ and $\frac{x_{ERR}}{x_o}(t_{ERR}) = 2\pi \frac{F_m}{F_0}\left|\frac{t_{ERR}}{T_0}\right|$ respectively. (The sidebar on

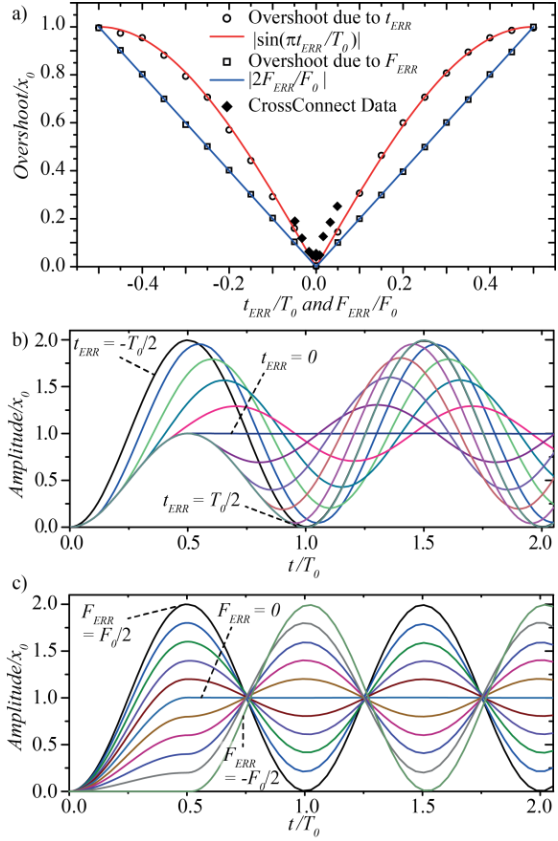

Figure 15. Overshoot resulting from drive errors. a) Overshoot as a function of timing and forcing errors obtained from finite element simulations (empty points) experiments of MEMS optical switches (solid diamonds) and analytic solutions based on Fourier transforms of the drive traces. b) and c) Amplitude response of a simulated MEMS with varying detuning parameters (even increments) in time and force respectively. The overshoot in a) is extracted from these plots.

the device itself resulting from its electrical and/or thermal load.

Many MEMS devices can be operated in the high *Q* limit where dissipative effects can be completely neglected, in which case only the frequency and electromechanical coupling need to be known. In practice one could envision a device where these two parameters are determined during self-calibration when the device is turned on. In a larger system, individual elements of an array of optical cross-connect mirrors could be periodically re-calibrated, while maintaining operational continuity of the system.

Feedforward and Feedback Drive Filters describes how the same results can be obtained using Laplace transformations.) This instability sets the limits to the overdrive actuation method. Considering overdrive $b$ and $F_m = 20F_0$: a 1% error in time results in a 126% overshoot and to limit the overshoot to 5%, the timing must be accurate to within 0.04% of the period of the resonator. Though surely not trivial, this has been successfully demonstrated in MEMS [31] where a timing accuracy of 200 ns is required. One should expect the resonant drive scheme to by highly sensitive to the experimental parameters: A high $Q$ device can no longer be actuated if the detuning in frequency is a considerable fraction of $\Delta f = f_0/Q$. Consequently, the acceptable uncertainty can quickly reach experimental limits, and any environmentally induced drifts would prove fatal to the approach. Stability, or robustness, to parameter changes of the system, can be improved arbitrarily by adding additional input commands [41], [69]. Zero vibration (ZV), equivalent to the double-step drive have been expanded to higher order zero vibration and derivative (ZVD) drive modes and beyond [43]. While such approaches allow parameter changes of over 30 % while still maintaining overshoots beneath 5%, it comes at a price; each additional command results in a delayed settling time. Where systems such as cranes need to account for large changes in mass, MEMS parameters are typically bound within a few percent of specs. Hence, the added stability of higher order feedforward terms is not required and the loss in response time becomes unjustified. Higher order pre-sculpting of the overdrive $b$ actuation scheme in MEMS did not lead to significant improvements in the settling times [31].

Compared to well-tuned feedforward drive schemes, closed-loop systems such as PID controllers cannot improve the settling time of well characterized and stable MEMS devices. Feedback systems are however efficient in eliminating external disturbances such as vibrations as well as drifts caused by environmental factors. To achieve the highest levels of stability a closed-loop system will always fare best. Feedforward drive sculpting can be combined with stabilizing feedback loops when both low response times and high stability/accuracy is required.

**The effects of imperfect drive sculpting: Smooth Forcing**

In all examples presented so far, it was assumed that the applied double-step force was ideal. This means that there is an infinitely sharp set of steps. The overdrive can be constructed from three step functions within very short intervals, occurring at $t = 0$ s, $t = t_1$ and finally at $t = t_2$. For any physical system, especially for electric drive schemes with long time constants, the approximation of a perfect step function may not be applicable. The stability and effect of forcing errors is discussed in [70]. To understand both how the multi-step drive works and the effect of non-idealized steps, one can calculate the Fourier transform of the drive force. This reveals the level of forcing as a function of frequency and is shown in Figure 16, where both the time and frequency domain of the single, double-step and overdrive forcing are plotted. (The analytical expressions of the Fourier transforms plotted are included in a the Frequency Domain Actuation Sidebar) In addition to the idealized steps, rounded steps are included by replacing the step functions with hyperbolic tangents, representative of smooth, more physical changes in the applied force. The amplitude of the Fourier transform of the step function has a simple analytic solution, falling off linearly without any features when plotted on a log-log curve. Consequently, all modes are actuated, although the higher frequency ones experience lower drive forces, as would be expected. The amplitude of the Fourier transform of the double-step drive differs in an important way: Superimposed over the linear fall off towards higher frequencies, there are periodic drops to vanishing forces at $f_0$, and higher harmonics. The double-step drive does not actuate the resonant mode because there is simply no force contribution at that frequency. If a system had higher order modes of the form $nf_0$ for odd $n$, these would also not be actuated. This also implies that very high frequency modes can be suppressed using low speed double-step forcing as long as $t_1=1/(nf_0)$.

It is interesting to note the effect of smoothing the transition. As previously stated, replacing the step with a smooth transition is the equivalent of adding a low-pass filter to the drive circuit. Consequently, the high frequency contributions are suppressed. The duration of the transition defines the maximal contributing frequency. For the examples plotted in Figure 16, it was assumed that the step took approximately $0.2T_0$ to complete. This results in a FFT with a sharp drop above roughly $2/T_0 = 2f_0$. The FFTs of the single and double-step drive modes both have the same underlying structure, the only difference being the periodically vanishing contributions at $nf_0$ for all odd $n$. Given the efficacy of the double-step drive, even with smoothed actuation amplitudes, it is possible to suppress higher order modes using a low-pass filter with a bandwidth up to $2f_0$. Experimentally, the falloff in the spectrum is visible in Figure 9 for the magnetic drive example. In that example, the drive coils act as a low-pass filter and effectively suppress signals above ~0.5 kHz.

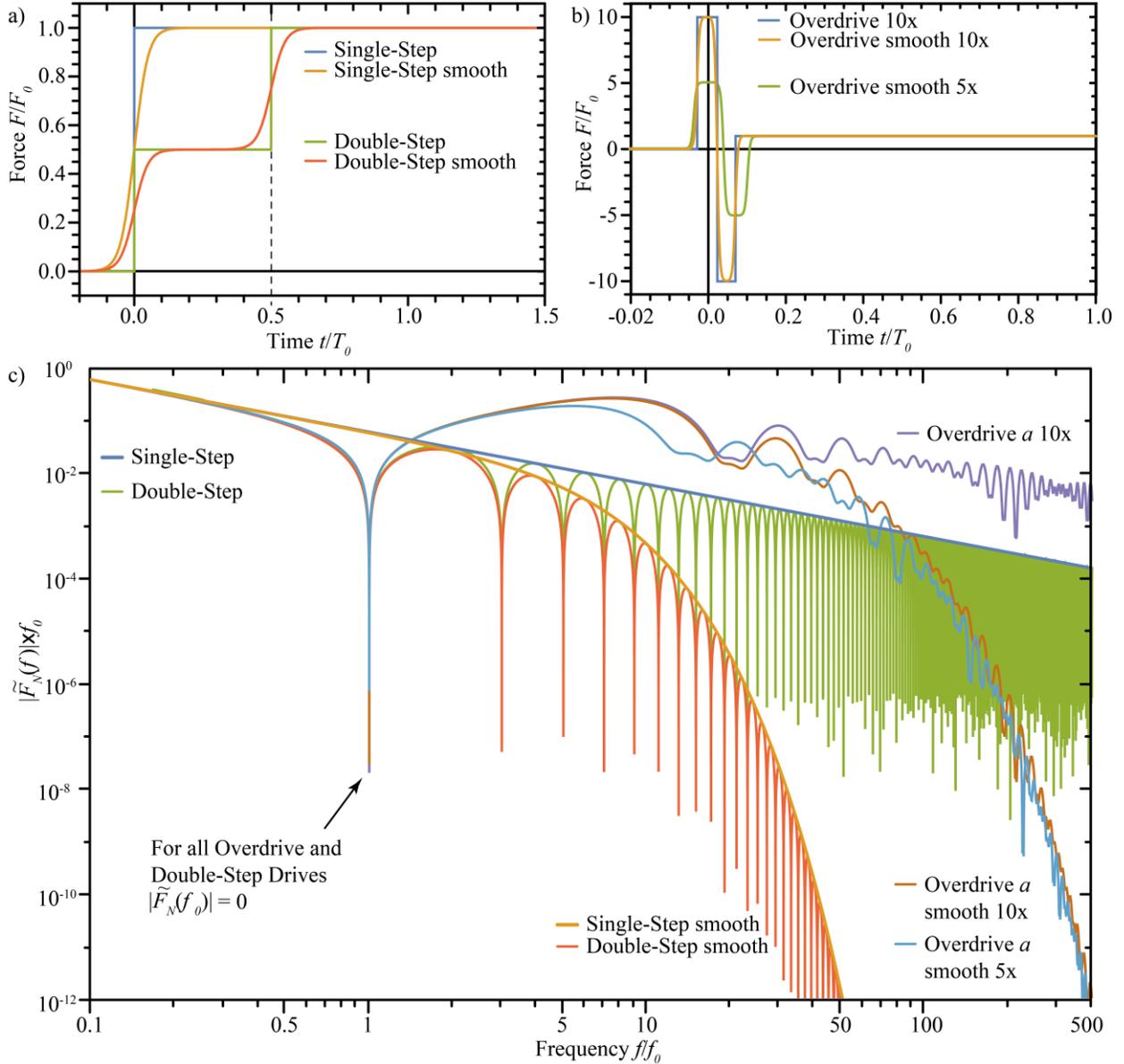

Figure 16. Actuation traces and their Fourier transforms. a) Single-step and double-step drive for a dissipation free resonator of period $T_0$. For each drive modality a smooth example is included with rounded edges, a closer approximation to experimental conditions which dictates that physical forces must be both continuous and smooth. b) Overdrive $a$ five- and tenfold of the final force. c) Fourier transform of actuation traces depicted in a) and b). For the idealized double-step drive and Overdrive, the force amplitude vanishes at $f_0$. The smooth drive forces fall off rapidly at higher frequencies, but have minimal deviation from the ideal drive at $f_0$. The frequency at which the amplitude falls off is related to how smooth, or rounded, the drive force is. The Overdrive results in high amplitudes and falls after a factor of 10 in frequency space compared to the double-step drive, in agreement with the ten times shorter timescales required to implement the drive scheme.

For the overdrive methods, this is not so simple. The short time intervals of $t_1$ and $t_2$ required when $F_m >> F_0$ prohibit the presence of a low pass filters. The FFT of the overdrive forcing plotted in Figure 16 also exhibits a dip at $f_0$, required to suppress on-resonance driving. However, unlike the double-step drive, the amplitudes at high frequencies also remain large. The three step functions defining the overdrive *a* produce three contributions to the $1/f$ spectrum. Furthermore, the amplitude scales with the step height, resulting in a larger prefactor to the $1/f$ dependency. Given that the time intervals are on the order of ten times faster, the resulting drive amplitudes are also still high at ten times the frequency. Even for the smoothed forcing scheme, the higher requirements on the response speed push the high frequency drop off to over $50f_0$. For comparison Figure 13 a) shows the overdrive *b* forcing traces and b) the corresponding Fourier transforms. Analogous to overdrive *a*, the shaped drives in frequency space have a vanishing contribution at $f = f_0$ (and odd harmonics). As the overdrive amplitude increases the drive approaches a delta function and the drive amplitude becomes flat in frequency space: Now all frequencies but $f_0$ are actuated.

All ringing can be suppressed by ramping the drive force with a controlled slope, effectively placing a low-pass filter with a cutoff below $f_0$. While this may be the simplest method technologically, the fastest possible ramp without ringing will settle a MEMS device on the order of $10T_0$, or 20 times slower than the ideal double-step drive.

## Conclusions

We have demonstrated analytically and experimentally that applying well-timed drive steps at the correct amplitude can completely eliminate the ringing of a damped (or undamped) resonator system. The simple point mass and spring model is shown to be directly applicable to complex MEMS devices. A three order of magnitude of reduction in the settling time for both capacitively and magnetically actuated mirrors is demonstrated experimentally. In a thermally driven system, the periodic transient deflections can also be effectively suppressed for which case drift behavior, resulting from long thermal relaxation times, dominates the settling time. These significant enhancements in performance are achieved without the need for self-sensing and closed-loop drive schemes. These techniques are presented as a method to tame high quality factor devices, simplifying the control electronics. The results demonstrate that an open-loop system perform point-to-point transitions at extremely high rates. Applications such as mechanically controlled optical switches will benefit considerably from implementing such drive modalities. Demonstrated by finite element simulations, it is suggested that properly engineered devices can be driven by high acceleration and deceleration forces, resulting in settling times of an additional order of magnitude shorter than already achieved using the double-step drive. It is believed that if such a drive modality were implemented in an actual device, its response would outperform active feedback circuits with the same maximal actuation force available.

The rapid point-to-point settling time relaxes some of the design constraints on MEMS devices. Consequently, the devices can be larger, resulting in enhanced optical characteristics, and the springs providing the restoring forces can be softer, thereby allowing for greater angular range. As an example, one may consider using extremely soft springs made of soft polymers to rotate a mirror over a large angle [71] and still achieve a useful response time. The effective cancellation of ringing could also be used for sensing. If an external perturbation shifts the resonant frequency (for example by adding mass), then the pre-programmed double-step drive would result in ringing, revealing the perturbation.

Given the proliferation of MEMS devices, it is believed that the a more comprehensive adoption of the feedforward command shaping will significantly enhance performance and enable new designs previously considered to be impractical.

MEMS devices are fabricated commercially to high tolerances and the mechanical properties do not vary significantly during the lifetime of the devices. Hence, feedforward in general, and the double-step drive in particular, provide an opportunity to significantly reduce the settling time without the need of complex closed-loop drive schemes. Faced with a challenging set of specifications the MEMS engineer can use these techniques to open up the design space and explore a more advantageous balance between size, performance and speed.

## Acknowledgements

The authors would like to thank CrossFiber, Inc. for their loan of the MEMS cross connect mirrors used for the studies present here. This work is funded by Boston University, LGS, BU Office of Technology and Development, Undergraduate Research Opportunities Program, DARPA Atoms to Product Program, NSF ERC, and the BU Photonics Center.

## Sidebar

## Open-Loop Step Response

The MEMS resonators considered here are guided by a few key assumptions. These include instantaneous force upon actuation and a linear response to the force. A linear, driven damped harmonic resonator is described by a differential equation of the form, $m\ddot{x} + \gamma\dot{x} + kx = F(t)$, where $m$ is the mass, $x$ is the amplitude of the oscillation, $\gamma$ is the damping factor, $k$ is the spring constant and $F(t)$ is the time dependent force. For an instantaneous step force, the differential equation can be written as

$$\ddot{x} + \frac{\gamma}{m}\dot{x} + \omega_0^2 x = \frac{F_0}{m} H(t_0), \quad (S1)$$

where $H(t_0)$ is the Heaviside function for a step function beginning at $t = t_0$ and $\omega_0 = \sqrt{\frac{k}{m}}$ is the natural frequency of the resonator. For all following equations $t_0 = 0$ can be assumed without any loss of generality.

The solution to equation (S1) is a superposition of a particular solution of the inhomogeneous equation and the general solution of the homogeneous equation (where $F_0 \equiv 0$). The particular solution depends only on the force function and is given by

$$x_p(t) = F_0/k. \quad (S2)$$

The homogeneous differential equation is linear with constant coefficients and can be solved by finding the roots of the characteristic equation. In the low dissipation limit ($\frac{\gamma}{2m\omega_0} < 1$) the roots are complex and $\omega_1 = \omega_0\sqrt{1 - (\gamma/(2m\omega_0))^2}$ is the resonance frequency of a damped harmonic oscillator. The general solution with two integration constants can be written as

$$x_g(t) = e^{-\frac{\gamma t}{2m}}(c_1 \sin \omega_1 t + c_2 \cos \omega_1 t). \quad (S3)$$

The solution is constrained by the initial conditions which require that $x(0) = 0$ and $\dot{x}(0) = 0$ for an unperturbed, stationary system at $t = 0$. Given the solution $x(t) = x_g(t) + x_p(t)$ the two coefficients $c_1$ and $c_2$ are determined. The zero amplitude condition, $x(0) = c_2 + F_0/k = 0$, requires that $c_2 = -F_0/(k)$. While the zero velocity at $t = 0$ s, $\dot{x}(0) = \frac{F_0}{k} \times \frac{\gamma}{2m} + c_1\omega_1 = 0$, requires that $c_1 = -\frac{F_0}{k} \times \frac{\gamma}{2m\omega_1}$. Reconstructing the total solution results in

$$x(t) = \frac{F_0}{k}\left(1 - e^{-\frac{\gamma t}{2m}}\left(\frac{\gamma}{2m\omega_1}\sin \omega_1 t + \cos \omega_1 t\right)\right). \quad (S4)$$

Redefining some of the variables in (S4) can simplify the equation significantly by making it dependent on only a single trigonometric function. Defining $\cos \phi = \gamma/(2m\omega_0)$ and keeping in mind the relationship between $\omega_0$ and $\omega_1$, the latter can be written as $\omega_1 = \omega_0 \sin \phi$. Rewriting (S3) in terms of a single trigonometric function with a phase $\phi$, the final form is given as

$$x(t) = \frac{F_0}{k}\left(1 - e^{-\frac{\gamma t}{2m}}\left(\frac{\cos \phi}{\sin \phi}\sin \omega_1 t + \cos \omega_1 t\right)\right), \quad (S5)$$

$$= \frac{F_0}{k}\left(1 - \frac{e^{-\frac{\gamma t}{2m}}}{\sin \phi}\sin\left(\sqrt{1 - \left(\frac{\gamma}{2m\omega_0}\right)^2}\omega_0 t + \phi\right)\right). \quad (S6)$$

*Ring-up for harmonic drive*

A harmonically driven resonator will respond as a function of the forcing amplitude and frequency. For on resonance drive, the amplitude is limited by the dissipation of the resonator. The general response of a system initially at rest takes the form

$$x(t) = A_t e^{-\frac{\gamma t}{2m}}\sin\left(\sqrt{1 - \left(\frac{\gamma}{2m\omega_0}\right)^2}\omega_0 t + \varphi_t\right) + A\cos(\omega t - \varphi). \quad (S7)$$

| Impedance [Ω] | $1/(\omega C)$ | $\omega L$ | $R(T)$ |
|---|---|---|---|
| Forcing Timescale $\tau$ [s] | $RC < 1$ ns | $L/R \sim 1$ ms | $\dfrac{L^2 \rho C_p}{\pi^2 \kappa_{th}} \sim 1$ ms |
| Force [N] / Torque [Nm] | $-\dfrac{1}{2}\dfrac{dC}{dx_i}V^2$ | $\mathbf{M} \times \mathbf{B}$ | $\sim E\alpha \Delta T A$ |
| Set point Power [W] | $\sim 0$ | $I^2 R$ | $I^2 R(T)$ or $V^2/R(T)$ |
| Transient Power (voltage biased [W] | $\dfrac{V^2}{R} e^{-\frac{t}{\tau}}$ | $\dfrac{V^2}{R}\left(1 - e^{-\frac{t}{\tau}}\right)$ | Dependent on $R(T)$, typically same order as the set point power |

Table S1: Summary of experimental parameters and expressions. $f_0$ and $Q$ are the resonance frequency and

**Sidebar**
**Experimental parameters**

|  | Electrostatic (CrossFiber) | Electrostatic (SLM) | Magnetic |  | Thermoelectric |
|---|---|---|---|---|---|
| $f_0$ [Hz] | 1657 | 80029 | 320, 1177 |  | 66.76 |
| $Q$ ($\frac{\gamma}{2m}$ [Hz]) | 265 (19.6) | 6.3 (37037) | 3100, 260 (0.323, 14.2) |  | 13.97 (15) |
| $t_1$ [s] ([$T_0$]) | 320×10⁻⁶ (0.536) | 6.6×10⁻⁶ (1.2) | 1.56×10⁻³, 0.42×10⁻³ |  | 7.5×10⁻³ (0.5) |
| $F_1$ [$F_0$] ([$P_0$]) | 0.52 | 0.59 | 0.5 |  | (0.531 up 0.520 down) |

With

$$A = \frac{F/m}{\sqrt{(\omega_0^2 - \omega^2)^2 + \left(\frac{\gamma\omega}{m}\right)^2}}, \quad (S8)$$

$$\varphi = \tan^{-1}\left(\frac{\gamma\omega}{k - m\omega^2}\right) - \varphi_d, \quad (S9)$$

$$\varphi_t = \tan^{-1}\left(\frac{\sqrt{1 - \left(\frac{\gamma}{2m\omega_0}\right)^2}\omega_0}{\left(\frac{\gamma}{2m} + \omega\tan\varphi\right)}\right), \quad (S10)$$

$$A_t = A\frac{\cos\varphi}{\sin\varphi_t}, \quad (S11)$$

where $\varphi_d$ is the phase of the drive and $\omega$ is the drive frequency. After the transient time has elapsed, the response becomes the well-known solution to the damped driven harmonic oscillator driven to amplitude $A$. In the example of harmonic drive given above, the drive frequency was set to resonance.

**Sidebar**

**Finite Element Simulations of resonant modes**

The finite element simulation program COMSOL was used to determine the mode shapes of the magnetically driven MEMS mirror depicted in Figure 2 b). The results are illustrated in Figure S1.

A single torsion device, like that used for the overdrive simulations depicted in Figure 12 a) has no other low frequency torsion modes. The next two higher modes related to a torque, or twisting of the springs are depicted in Figure S2. These results demonstrate that the overdrive actuation scheme may be applicable to a purely torsional system, lacking asymmetries as those of the magnetically driven system described above.

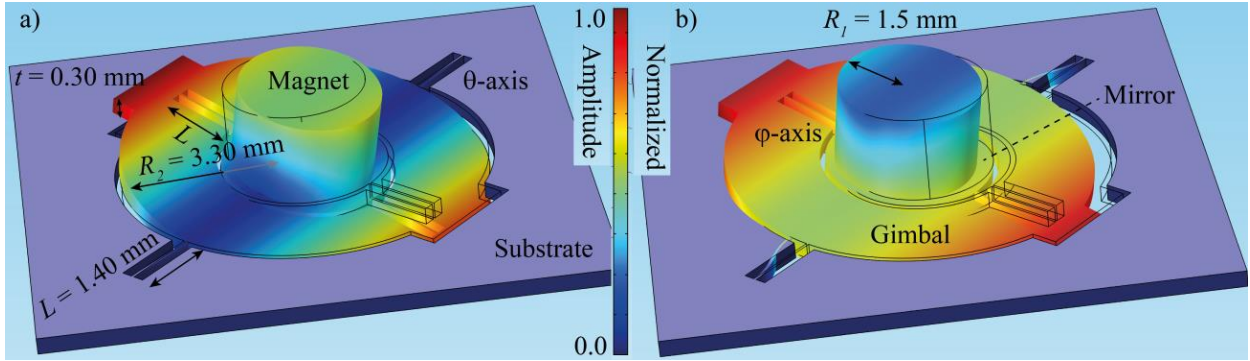

Figure S1. Finite element simulation of a dual axis mirror with a magnet attached. Color indicates normalized displacement. a) The fundamental torsional mode was found experimentally at 320 Hz. b) The higher order mode, also actuated by the externally applied torque, experimentally found at 1179 Hz. This mode is the planar flexural mode, where the tilt results from the asymmetry caused by the magnet. These two modes couple when a step in the force is applied.

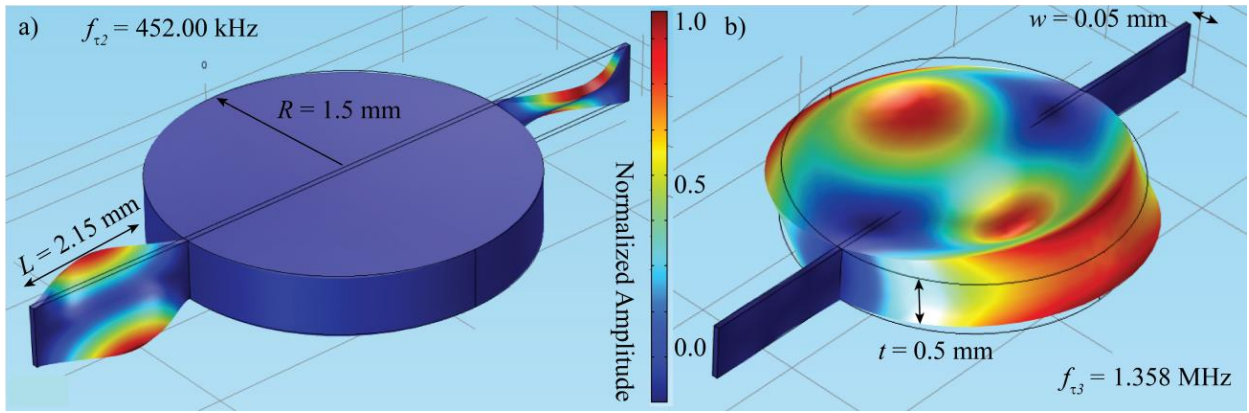

Figure S2. Finite element simulation of single axis mirror. The higher order torsion modes occur at a) $f_{\tau 2}$ = 452 kHz and b) $f_{\tau 3}$ = 1.3558 MHz, where the fundamental torsion mode (Fig. 15a)) has a resonance frequency of $f_{\tau 1}$ = 320 Hz.

## Sidebar
## Thermal Relaxation and Piezoresistivity

Thermal relaxation measurements using a constant voltage bias are presented. As the power is proportional to $V^2/R$ and $R(T)$ is a monotonically increasing function in the range considered, there is an initial power spike, or excess, before the system finds its thermal equilibrium. This is illustrated by the measurements shown in Figure S3. The overshoot of approximately 10% shows an exponential decay to the equilibrium position governed by two thermal time constants. The shorter timescale of 1.02 ms is given by the thermalization time of the bimorphs, while the longer timescale of 39.8 ms is the thermalization time of the whole system, including the large mirror which acts as a heat sink and the silicon base itself on which the MEMS device is situated. The oscillations visible in the thermal measurements shown in Figure S3 are the result of piezoresistive response of the bimorphs, where the resistance is modulated by the mechanical strain as described by equation S12. For the thermal drive to be able to cause mechanical ringing, the thermal timescales must be faster than the mechanical timescale. Given a mechanical period of 15.0 ms, the thermal timescale of just over one millisecond is sufficiently fast for both resonant actuation as well as the double-step drive which requires the duration of the first drive power to be 7.5 ms, during which the system can find its thermal equilibrium.

A piezoresistive material exhibits a strain sensitive resistivity described by

$$\gamma_P = \frac{1}{\varepsilon_s}\frac{\Delta R}{R} = (1+v) + \frac{1}{\varepsilon_s}\frac{\partial \rho}{\rho}, \quad (S12)$$

where $\rho$ is the resistivity, $v$ is the Poisson's ratio, and $\varepsilon_s$ is the mechanical strain. $\gamma_P$, known as the gauge factor, is a measure of the strength of the piezoresistive effect [65].

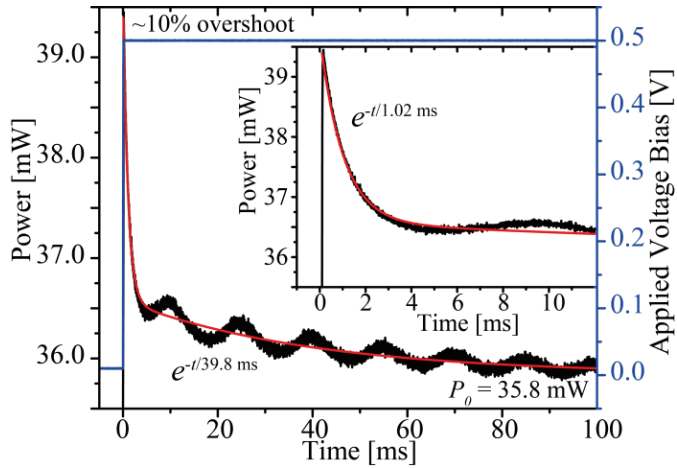

Figure S3. Thermal relaxation measurements. Measuring the current after a voltage step reveals that the thermal relaxation of the device is governed by two time scales, that of the bimorphs and that of the entire substrate-bimorph-mirror system. The observed ringing has the same period as the mechanical response and results from a strain dependency on the resistance (piezoresistivity).

## Sidebar

## Feedforward and Feedback Drive Filters

*Feedforward filters*

In control theory Laplace space is used to determine the command required to obtain a desired outcome. An overview of control theory is given by [50]. The block diagrams in figure S4 depict different scenarios. In figure S4 a) the simplest case is illustrated, where only a feedforward command is included. In this formalism the response is expressed as

$$Y(s) = R(s)F(s)G(s), \quad (S13)$$

where the position is determined by taking the inverse Laplace transformation of $Y(s)$ as $y(t) = L^{-1}[Y(s)]$. Here $R(s)=1/s$ represents the step function in Laplace space and $G(s)$ is the second order transfer function describing an underdamped harmonic oscillator and which takes the form

$$G(s) = \frac{\omega_0^2}{s^2 + 2\zeta\omega_0 s + \omega_0^2}, \quad (S14)$$

with $\zeta = \frac{\gamma}{2m\omega_0}$.

The input command, or command shaping in Laplace space is $T(s) = R(s)F(s)$. As R(s) is given by the controller (here the desired response is a step function) the task remains to define the filter $F(s)$. This can be achieved by applying the Posicast Input Command Shaping method (PICS) and is illustrated in the literature for general transfer functions [44], [72] and for resonant systems in particular [51]. In this article the 'filter' was derived in time domain. Transforming the results into Laplace space reveals the prefilter formalism used in control theory.

Consider the double-step drive: The input command is a two stage step function characterized by $F_1$ and $t_1$. Hence, in Laplace space the command becomes

$$T(s) = L[F_1 H(0) + (1-F_1)H(t_1)] = \frac{1}{s}(1 + (1-F_1)e^{-t_t s}), \quad (S15)$$

where $H(t)$ is again the Heaviside function. As $R(s)=1/s$ one now knows the input filter in Laplace space:

$$F_{DS}(s) = (1 + (1-F_1)e^{-t_1 s}). \quad (S16)$$

One of the strengths of this method is that it can be applied to each mode of the resonator in a general fashion. Like this the two fold double-step has a input filter for the form $F_{DS1}(s)F_{DS2}(s) = (1 + (1-F_1)e^{-t_1 s}) \times (1 + (1-F_2)e^{-t_3 s})$. It must be remembered that the second order transfer function must also be modified so that $G(s) = G(s,\omega_0,\zeta_0) \times G(s,\omega_1,\zeta_1)$. It naturally follows that this can be expanded to include an arbitrary number of modes.

Analogously, one can determine the input filter for the overdrive *a* and *b* actuation schemes, to obtain

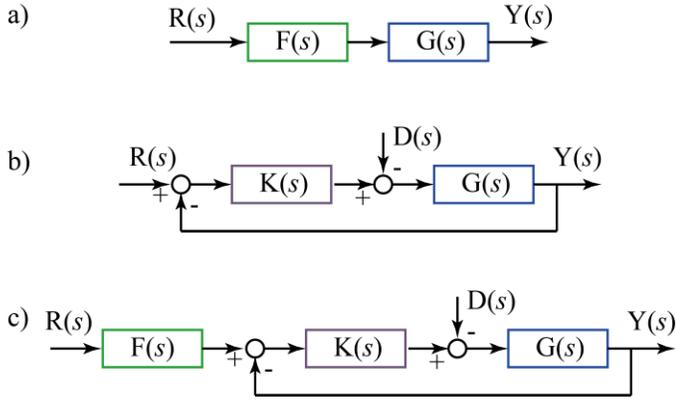

Figure S4. Control theory diagrams. a) Feedforward only command. b) Feedback, K(s) is based on a PID control mechanism, c) implementation of feedforwards combined with feedback. D(s) represents a disturbance, if it is known (measured) then it may also be compensated using feedforward command shaping.

$$F_{Overdrive\ a}(s) = (F_m - 2F_m e^{-t_1 s} + (1 + F_m)e^{-t_2 s}) \quad (S17)$$

$$F_{Overdrive\ b}(s) = (F_m - F_m e^{-t_1 s} + e^{-t_2 s}). \quad (S18)$$

The smoothed input commands do not have simple closed form solutions but the same method can be applied numerically as needed.

Using the prefilter formalism a stability analysis can be performed, where the result reproduces the solution discussed in the main text which used the Fourier transform and simulation methods to determine the sensitivity to detuning. The numerical solutions to the Laplace transformation can be determined more easily than the simulations and is well suited for such studies. Using this method it was confirmed that the overshoot scales as $\Delta x_{DS} = \frac{\pi}{F_0}\left|\frac{t_{ERR}}{T_0}\right|$, $\Delta x_{Overdrive\ a} = \frac{4\pi F_m}{F_0}\left|\frac{t_{ERR}}{T_0}\right|$, and $\Delta x_{Overdrive\ b} = \frac{2\pi F_m}{F_0}\left|\frac{t_{ERR}}{T_0}\right|$, with respect to small detuning in $t_1$, $t_{1a}$ and $t_{1b}$ respectively.

*Feedback Control using PID*

For comparison with the feedforward input filters, it is illustrative to determine what a feedback drive scheme may look like. The PID loop is the most commonly implemented feedback method, typically used for flow rate or temperature control systems. For a second order system the ideal P, I and D parameters can be determined analytically. The PID function transforms the second order transfer function into a first order transfer function, which exponentially approaches the desired set-point. In theory there is no minimum in the obtainable response time, however, experimentally instabilities and maximal applicable forces impose lower bounds. The model illustrated in figure S4 b) in equation format becomes

$$Y(s) = K(s)G(s)(R(s) - Y(s)). \quad (S20)$$

The transfer function is the ratio of the output over the input and hence

$$T(s) = \frac{Y(s)}{R(s)} = \frac{1}{\frac{1}{K(s)G(s)}+1} = \frac{1}{\tau s+1}, \quad (S21)$$

which is a first order transfer function with a well behaved response. Following the notation of [73] one can determine an ideal transfer function of the form

$$K(s) = \frac{K_p T_d}{s}\left(s^2 + \frac{1}{T_d}s + \frac{1}{T_i T_d}\right), \quad (S22)$$

which eliminate the poles for carefully chosen $T_d$ and $T_i$ parameters. This is achieved by setting $T_d = \frac{1}{2\zeta\omega_0}$ and $T_i = \frac{2\zeta}{\omega_0}$. Plugging (S22) and (S14) into (S21) it is possible to solve for the time constant $\tau = \frac{2\zeta}{\omega_0 K_p}$. Typically the settling time is defined when the approach as reached within 2% of the target or $t_s = 4\tau$.

In principle the PID feedback loop can be tuned to set the response time arbitrarily small. $K_p$ however is divergent for washing $\tau$, corresponding to an unbounded force, which is of course prohibited. Furthermore, instabilities make the implementation of PID loops impossible for very high gains. Lastly, while it can be shown that a PID controlled resonator can settle by $T_0/2$ while maintaining a force $\leq F_0$, this is only achieved by carefully tuning the control loop parameters which again depend on $f_0$ and $Q$. As a result the same knowledge of the system is required as was used for the feedforward drive: Drifts in the MEMS parameters will also set a limit to the effectiveness of feedback loops. Just as observed for the overdrive examples, the overshoot due to detuning errors scales with the required forcing.

For implementation the feedback poses additional burdens. Fabrication needs to include sensitive sensing mechanism, increasing complexity and often adding to the MEMS size and cost while degrading performance. Finally, during operation continuous feedback would be needed. Finite sampling time [31] and minimum measurement integration time can also set a limit to the efficiency of a feedback loop.

This is not to claim feedback is not without its advantages. It is noted in [73] that the primary capacity of a PID feedback system is to eliminate offset (or steady state error). For MEMS this means that a PID loop can effectively eliminate drift, it is not ideally suited to eliminate the resonant response which can be done more efficiently with a much simpler implemented feedforward command. A possible ideal implementation is illustrated in figures S4 c), which combines both feed forward and closed loops drive sculpting, and is discussed fully in the literature [50]. Such control systems are can also mitigate a disturbance factor which, if measured, can be compensated using feed forward filters.

### Sidebar

### Drive Force in Frequency Domain

This Sidebar presents the analytical Fourier transforms of the drive signals for the single-step, double-step and overdrive forcing plotted in Figures 13 and 14.

Fourier Transform of single-step drive

$$FSSD(f) = \frac{F_0}{2\sqrt{2}}\left(\frac{i}{\pi^{\frac{3}{2}}f} + \frac{\delta(f)}{\sqrt{\pi}}\right), \quad (S23)$$

where $\delta(f)$ is the delta function. The corresponding magnitude for $f > 0$ Hz:

$$|FSSD(f)| = \frac{F_0}{2\sqrt{2}\pi^{\frac{3}{2}}}\frac{1}{f}, \quad (S24)$$

As described, the amplitude scales with the drive force and falls of as $1/f$ as a function of frequency, suppressing higher order mode actuation.

Fourier Transform of double-step drive

$$FDSD(f) = \frac{F_0}{2\sqrt{2}}\left(\frac{i(1+e^{i\pi T_0 f})}{2\pi^{\frac{3}{2}}f} + \frac{\delta(f)}{\sqrt{\pi}}\right). \quad (S25)$$

And the corresponding magnitude for $f > 0$ Hz

$$|FDSD(f)| = \frac{F_0}{4\sqrt{2}\pi^{\frac{3}{2}}}\frac{\cos(\frac{\pi}{2}T_0 f)}{f}, \quad (S26)$$

For the double-step drive, in addition to the $1/f$ fall off there is a cosine modulation, ensuring that the drive vanishes as $\cos\left(\frac{\pi}{2}f\right) = 0$ or for $f = nT_0$, for all odd $n$.

Fourier Transform of over-drive

$$FODa(f) = \frac{1}{2\sqrt{2}}\left(\frac{i\left(F_m(1-2e^{i2\pi t_1 f})+e^{i\pi t_2 f}(F_m+F_0)\right)}{\pi^{\frac{3}{2}}f} + \frac{\delta(f)}{\sqrt{\pi}}\right). \quad (S27)$$

With no analytical solution for the corresponding magnitude.

Fourier Transform of over-drive with no externally applied deceleration

$$FODb(f) = \frac{1}{2\sqrt{2}}\left(\frac{i\left(1-e^{i\,2\pi f\,t_1}+e^{i\,2\pi f\,t_2\frac{F_0}{F_m}}\right)}{\pi^{\frac{3}{2}}f\frac{F_0}{F_m}} + \frac{\delta(f)}{\sqrt{\pi}}\right). \quad (S28)$$

With no analytical solution for the corresponding magnitude.

## AUTHOR INFORMATION

*Matthias Imboden* is currently a Post-Doctoral Fellow at EPFL working on soft transducers for biomechanical applications. He received his Diploma at Bern University in 1999 and completed his Ph.D. dissertation on diamond NEMS at cryogenic temperatures at the Physics Department of Boston University. His expertise is in the study of NEMS and MEMS and their application in physics research. Academic interests include novel lithography techniques, non-linear mechanics, superconductivity, electro-mechanical transduction of the Casimir effect, and biomechanics. He is a member of the American Physical Society.

*Jackson Chang* received the M.Eng. in Materials Science and Engineering from Boston University in 2012. He currently works as a Research Engineer under Dr. David Bishop. His research focuses on large out-of-plane actuation MEMS devices and their applications in RF technology. Topics of interest include RF MEMS, MOEMS, and microfabrication.

*Corey Pollock* received his B.S. in Mechanical Engineering from the University of Washington in 2010. He is currently pursuing a Ph.D. in Mechanical Engineering at Boston University as a member of David Bishop's group. His research involves MEMS technology, specifically studying micro mirrors and smart lighting applications.

*Evan Lowell* is currently a senior studying Mechanical Engineering at Boston University. His research background is focused on MEMS devices. He won the Boston University Imagineering Design Competition in Spring 2015. Mr. Lowell is a member of the Material Research Society as well as the American Society of Mechanical Engineers. He also is a Dean Lutchen fellow and part of the Brawerman Leadership Fellowship.

*Mehmet Akbulut* is currently a Mechanical Engineering undergraduate student at Boston University. Mehmet worked at SpaceX for the In-Space Propulsion Group during 2015. He won the Imagineering Competition in 2015 for developing and demonstrating a fully automated robotic guitar player. His expertise is in the design, control and analysis of electro-mechanical systems. Academic interests include aerospace, robotics and sustainable energy. Mehmet is currently the president of American Institute of Aeronautics and Astronautics branch at Boston University. Mehmet was recently elected to sit on the national Board of Directors for SEDS-USA, a 501(c)3 non-profit organization established in 1982 to promote STEM education and provide students with opportunities in the aerospace field.

*Jessica Morrison* received a B.S. in physics from the University of Cincinnati in 2011. She began graduate coursework at Boston University in physics in 2011. She is currently working as a research associate at Boston University under Dr. David Bishop. Her research involves the use of optical microelectromechanical systems in smart lighting systems and nonlinear mechanical effects in large range optical MEMS. She is a member of the American Physical Society, the Materials Research Society, and the Optical Society of America.

*Thomas Stark* received his B.S. in Physics from Middlebury College in 2011, where he conducted research on laser cooling rubidium. He is currently pursuing his Ph.D. in Materials Science and Engineering at Boston University, where he is a member of David Bishop's group. His research focuses on using MEMS for fabricating tunable metamaterials. He is a member of the Materials Research Society and the American Physical Society.

*Thomas G. Bifano* directs the Boston University Photonics Center comprised of forty-five faculty members from seven academic departments with active programs for education, scholarly research and development of advanced photonic device prototypes for commercial and military applications. He is a Professor of Mechanical Engineering, and was Chair of the Manufacturing Engineering Department from 1999-2006. He served on the US Army Science Board from 2010-2014, and is a founder and Chief Technical Officer of Boston Micromachines Corporation in Cambridge MA. He earned B.S. (1980) and M.S. (1983) degrees in Mechanical Engineering and Materials Science from Duke University, and a Ph.D. (1988) in Mechanical Engineering from North Carolina State University. His research focuses on design and manufacturing of microelectromechanical systems (MEMS) for optical applications, especially development of wavefront control technology used in high-resolution telescopes, microscopes, and bioimaging systems. He was awarded the Bepi Columbo Prize in 2009 for his achievements in research, education, and technology development related to deformable mirror development and space science.

*David J. Bishop* is currently the Head of the Division of Materials Science and Engineering at Boston University (BU). He is also a Professor of Physics, Professor of Electrical and Computer Engineering, a Professor of Mechanical Engineering and a Professor of Materials Science and Engineering at BU. Previously he was the Interim Associate Dean for Research and Graduate Programs for the BU College of Engineering. Prior to joining BU, he was the Chief Technology Officer and Chief Operating Officer of LGS, the wholly-owned subsidiary of Alcatel-Lucent dedicated to serving the U.S. federal government market with advanced R&D solutions. Before joining LGS, Dr. Bishop was the President of Government Research & Security Solutions for Bell Labs, Lucent Technologies. He is a Bell Labs Fellow and in his previous positions with Lucent served as Nanotechnology Research VP for Bell Labs, Lucent Technologies; President of the New Jersey Nanotechnology Consortium and the Physical Sciences Research VP. He joined AT&T-Bell Laboratories Bell Labs in 1978 as a postdoctoral member of staff and in 1979 became a Member of the Technical Staff. In 1988 he was made a Distinguished Member of the Technical Staff and later that same year was promoted to Department Head, Bell Laboratories. Dr. Bishop graduated from Syracuse University with a B.S. in Physics. He received an M.S. and Ph.D. in Physics from Cornell University.